\documentclass[11pt]{article}
\usepackage[a4paper, left=26mm, right=26mm, top=25mm, bottom=30mm]{geometry}
\usepackage{graphicx} 
\usepackage{subcaption}
\usepackage{mathtools}
\usepackage{diagbox}
\usepackage{amssymb}
\usepackage{physics}
\usepackage{float}
\usepackage{standalone}
\usepackage{csquotes}
\usepackage{enumitem}
\usepackage{todonotes}
\usepackage[normalem]{ulem}
\usepackage{xurl} 

\usepackage[hidelinks]{hyperref}
\usepackage{bbold}
\usepackage{cite}

\usepackage[table]{xcolor}

\graphicspath{{figs/}}

\begin{document}

\begin{titlepage}
    
\begin{center}
${}$\\
\vspace{100pt}
{ \Large \bf Quantum Gravity and Effective Topology
}

\vspace{36pt}

{\sl J. van der Duin}$\,^{\dagger}$, {\sl R.\ Loll}$\,^{\dagger,\star}$, {\sl M. Schiffer}$\,^{\dagger}$ and {\sl A. Silva}$\,^{\dagger}$

\vspace{18pt}
{\footnotesize

$^\dagger$~Institute for Mathematics, Astrophysics and Particle Physics, Radboud University \\ 
Heyendaalseweg 135, 6525 AJ Nijmegen, The Netherlands.\\ 

\vspace{5pt}
{\it and}\\
\vspace{5pt}

$^\star$~Perimeter Institute for Theoretical Physics,\\
31 Caroline St N, Waterloo, Ontario N2L 2Y5, Canada.\\
}
\vspace{48pt}

\end{center}

\begin{center}
{\bf Abstract}
\end{center}

\noindent 
We introduce a new methodology to characterize properties of quantum spacetime in a strongly quantum-fluctuating regime, using tools from topological data analysis. Starting from a microscopic quantum geometry, generated nonperturbatively in terms of dynamical triangulations (DT), we compute the Betti numbers of a sequence of coarse-grained versions of the geometry as a function of the coarse-graining scale, yielding a characteristic ``topological finger print". We successfully implement this methodology in Lorentzian and Euclidean 2D quantum gravity, defined via lattice quantum gravity based on causal and Euclidean DT, yielding different results. 
Effective topology also enables us to formulate necessary conditions for the recovery of spacetime symmetries in a classical limit.

\vspace{12pt}
\noindent

\end{titlepage}

\section{Lattice quantum gravity and observables}
\label{sec:intro}

The advent of powerful, quantum field-theoretic lattice methods that take both the dynamical and the Lorentzian nature of four-dimensional 
spacetime into account makes
the challenge of understanding quantum gravity nonperturbatively much more concrete and well defined \cite{review1,review2,ency}. 
At an abstract level, the task resembles that of lattice QCD, namely to evaluate the nonperturbative path integral
in a suitable scaling limit and for an interesting range of physical scales. 
The gravitational path integral, also called ``sum over histories", is given by the formal functional integral 
\begin{equation}
Z=\int {\cal D}[g]\ \mathrm{e}^{\, i S^{\mathrm{EH}}[g]}
\label{pathint}
\end{equation}
over diffeomorphism equivalence classes $[g_{\mu\nu}]$ of spacetime metrics $g_{\mu\nu}$, where
\begin{equation}
 \;\;\;\; S^{\mathrm{EH}}[g]= \frac{1}{16\pi G_{\rm N}}\, \int_M d^4 x\, \sqrt{|\det(g)|}\, (R -2 \Lambda )
\label{seh}
\end{equation}
is the Einstein-Hilbert action with a cosmological term.
However, because of the very different field content and symmetry structure of general relativity and nonabelian gauge field theory, 
the technical requirements of the lattice set-up and the nature of the invariant quantum observables differ substantially.

Modern lattice quantum gravity in terms of causal dynamical triangulations (CDT) comes with a computational framework based on Markov chain Monte Carlo (MCMC) methods, 
adapted to gravity. It currently allows for the investigation of system sizes of about $10^6$ building blocks and for measuring observables in a near-Planckian scale window, a rare asset in quantum gravity. CDT lattice quantum gravity combines three key structural features that have enabled its breakthrough results\footnote{They include the nonperturbative emergence of a quantum spacetime with de Sitter features \cite{Ambjorn2004,Ambjorn2007,qrc3} and the discovery of dynamical dimensional reduction \cite{Ambjorn2005}, see also \cite{review1,review2,ency} for reviews.}: (i) the use of \textit{dynamical} instead of fixed hypercubic lattices, reflecting the dynamical nature of spacetime geometry, (ii) the \textit{exact} implementation of relabelling symmetry, the lattice analogue of diffeomorphism symmetry, and (iii) the presence of a Wick rotation for curved lattice spacetime configurations, which has no counterpart in the continuum and unlocks the application of MCMC technology. 

Being able to evaluate the expectation values
\begin{equation}
\langle {\cal O} \rangle = \frac{1}{Z} \int {\cal D}[g]\, {\cal O}[g]\, \mathrm{e}^{\, i S^{\mathrm{EH}}[g]},
\label{obs}
\end{equation}
of geometric observables $\cal O$ in the deep UV regime is a game changer for quantum gravity \cite{Loll2025}: it enables quantitative reality checks in a realm where, as a rule, classical geometric intuition fails due to the presence of large quantum fluctuations of spacetime itself. Importantly, these fluctuations prevent the existence of structures that could meaningfully serve as (quasi-)classical local reference frames. This also 
implies that most of the local tensorial constructions of textbook general relativity have no obvious, well-defined quantum counterparts \textit{as a matter of principle}.\footnote{This goes beyond the need to regularize and renormalize, which continues to hold, like in nongravitational relativistic quantum field theories.} 
Nevertheless, even without a smooth structure, length and volume measurements are still available, and serve as elementary ingredients in the construction of geometric observables. Lastly, the absence of any \textit{a priori preferred} background structure in pure gravity requires observables to be nonlocal, as may be achieved e.g.\ through spacetime averaging. 
These characteristics of observables indicate the kind of information one can gain from a fully-fledged quantum gravity theory, and reflect the physical nature of what constitutes ``quantum spacetime" in a nonperturbative realm.

This sets the stage for the present work, where we elaborate on a new class of observables introduced in \cite{letter} as a new tool to characterize the microscopic properties of quantum geometry. 
Generally speaking, observables have been investigated using CDT lattice methods with two main motivations: 
firstly, to provide tests of the classical limit, by showing that 
their eigenvalues on sufficiently large scales are compatible with (semi-)classical expectations. Successful examples are the \textit{shape} of the universe, i.e.\ its spatial three-volume $V_3$ as a function of proper time $\tau$, together with its quantum fluctuations $\delta V_3(\tau)$ \cite{Ambjorn2007,Ambjorn2008}, 
and its average scalar curvature \cite{qrc3}, which all turn out to match those of a de Sitter space. A second objective is to uncover new physics, in the form of genuine quantum signatures. The corresponding observables must of course be diffeomorphism-invariant and operationally well defined in the lattice framework, but by virtue of their nonperturbative nature may not relate in any straightforward way to the geometric properties of classical spacetimes. 
These observables can take the form of scaling exponents, characterizing universal aspects of quantum spacetime in a Planckian regime.  An example is the anomalous behaviour of the \textit{spectral dimension}, which in CDT lattice quantum gravity was discovered to be 2 (within error bars) near the Planck scale, instead of the classically expected value of 4, in what has since been conjectured to be a universal property of quantum gravity \cite{Carlip2017}. 

In a companion paper \cite{letter} we highlight this second motivation, using effective topology as a tool to quantitatively assess the ``foaminess" of quantum geometry near the Planck scale, which is often conjectured to be some kind of \textit{quantum spacetime foam} \cite{Carlip2022}. In the present work, we want to emphasize a motivation that relates to the classical limit, more precisely, the recovery of global, cosmological symmetries. This is particularly suggestive in view of the global de Sitter-like properties that have already been found in four dimensions and the fact that classical de Sitter space has a maximal number of isometries.
It is not obvious that symmetries are or should be present in a Planckian regime, or can even be defined there in a operationally meaningful way. 
One may nevertheless expect that \textit{at a sufficiently coarse-grained scale} some notion of \textit{approximate} symmetry may apply. As discussed in
\cite{Loll2023} and supported by a proof of principle, this requires the construction of diffeomorphism-invariant measures of homogeneity and isotropy, which is highly nontrivial. 

Rather than developing this idea further, we want to point out here that the presence of symmetry, in the sense of an (approximate) invariance under a suitable notion of continuous translation or rotation, 
needs a suitable carrier space on which such transformations can be defined. This implies that at the coarse-graining scale considered, a quantum spacetime must be sufficiently similar to (a piece of) $\mathbb{R}^4$ in topological terms, without any
topological obstructions in the form of holes, nontrivial connectivity (think wormholes) or parts of spacetime that become effectively disconnected. 
Investigating the effective topology of quantum geometry, as we will do below, allows us to formulate necessary criteria for being ``sufficiently nice" to support notions of symmetry at a given scale. As we will see, already for the two-dimensional toy models of Lorentzian and Euclidean quantum gravity, the outcomes are very different.  

The remainder of this paper is organized as follows. 
We introduce the concept of effective topology in Sec.\ \ref{sec:eff}, and explain in Sec.\ \ref{sec:hom} how methods from topological data analysis (TDA) can help in investigating observables related to this concept in the context of lattice quantum gravity based on dynamical triangulations. 
Sec.\ \ref{sec:coarse} contains a detailed description of how one proceeds in the particular case of two dimensions to generate a coarse-grained triangulation
whose homology is then measured.  This requires several steps: 
the selection of a coarse-grained vertex sample (Sec.\ \ref{sec:vertex}), the construction of the associated Voronoi decomposition (Sec.\ \ref{sec:voronoi})
and subsequently its dual, coarse-grained Delaunay triangulation (Sec.\ \ref{sec:delaunay}). In Sec.\ \ref{sec:effhom} we describe and discuss the results of numerically measuring
the expectation values of the Betti numbers for 2D Lorentzian and 2D Euclidean quantum gravity. Our conclusions and an outlook are contained in Sec.\ \ref{sec:concl}.
Appendix A contains additional technical details on how to construct the Voronoi decomposition, and Appendix B describes how to locally adjust 
Delaunay triangulations to make them amenable to the computational library used to determine the Betti numbers.

\section{Introducing effective topology}
\label{sec:eff}

The key idea behind the new observables is to examine the connectivity properties of the quantum geometry -- what we will call its 
\textit{effective topology} or, more precisely, its \textit{effective homology} -- on small scales relative to their total linear extension, 
as a function of a linear coarse-graining scale $\delta$. 
Recall that the homology of a
topological space keeps track of the number of its ``holes" of various dimensions, where the dimension is defined as that of the hole's boundary.\footnote{Note that homology is also characterized by torsion coefficients. They are not computed by the open-source library we use, and will not be considered in this work.} For example,
removing a two-dimensional disc from a two-dimensional sphere creates a one-dimensional hole, since the hole's boundary is a one-dimensional circle. 

Note that the topology of the spacetime configurations contributing to the lattice path integral is always fixed, usually to that of a sphere or torus, 
and not allowed to change during the Monte Carlo evolution.\footnote{The gravitational path integral cannot be renormalized (unambiguously) by standard methods if a sum over topologies is included, due
to the superexponential growth of the number of configurations as a function of the spacetime volume (see e.g.\ \cite{Loll2022}, Q14, for a discussion). The spaces of fixed topology we consider here
are always compact without boundary.} However, due to the nonperturbative character
of the gravitational dynamics, typical configurations in the continuum limit are highly nonclassical and nowhere differentiable, analogous to the 
nonclassical paths that support the Wiener measure of the quantum-mechanical path integral for a nonrelativistic particle. In particular, the microscopic
building blocks can arrange themselves into geometries that are macroscopically indistinguishable from spaces with a \textit{different} topology. 

\begin{figure}[t]
\centering
\includegraphics[width=0.55\textwidth]{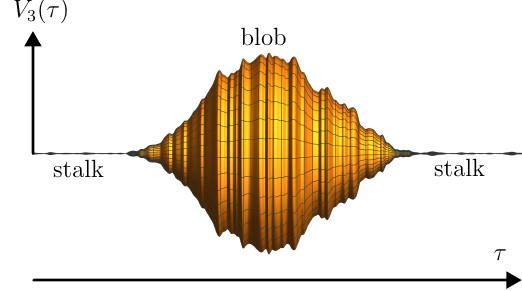}
\caption{Shape of a typical path integral configuration in 4D CDT lattice quantum gravity, illustrating the macroscopically emergent
$S^4$-topology. The curve $V_3(\tau)$ has been made into a rotational body about the horizontal time axis $\tau$.
}
\label{fig:blob}
\end{figure}

An example, in this case involving the global topology of spacetime, has been observed in 4D quantum gravity. 
The most common choice of lattice topology for this system is $S^1\!\times\! S^3$, where for technical convenience the time direction 
has been compactified 
to a circle.\footnote{4D CDT has also been investigated on a four-torus $T^4$. A discussion of how the choice of global topology may influence 
the nature of phase transitions can be found in \cite{Ambjorn2022}.} 
However, it turns out that in the so-called de Sitter phase the overall shape of the quantum geometry (the system's nonperturbative ground state) 
is driven dynamically to that of a four-sphere $S^4$ \cite{Ambjorn2004}! More specifically, a typical member of the ensemble of
geometries consists of
a thin ``stalk", a proper-time interval during which the spatial three-volume $V_3(\tau)$ is close to the kinematically allowed minimum, whose spatial extension
is of the order of the UV cut-off $a$ (the length of a lattice edge), and a spherical ``blob", a complementary time interval during which the spatial universe is macroscopically 
extended, $V_3(\tau)\gg a^3$ (Fig.\ \ref{fig:blob}). From a macroscopic point of view (where $a\rightarrow\! 0$ in a continuum limit) 
the stalk has vanishing volume and can be neglected relative to the total volume of the blob, leading to an``effective" large-scale topology $S^4$.
  
We will look systematically for the possible presence of nontrivial effective topology at scale $\delta$ (with $\delta$ assumed small compared 
to the total system size), as a novel way to characterize the local nature of quantum geometry. 
Roughly speaking, this is meant to capture the topology ``felt" by a probe of linear extension $\delta$ or a wavelike excitation of wave length $\delta$, and is
related to the presence of holes of characteristic linear size $\leq\delta$. To determine the effective topology in practice, we apply a single coarse-graining step of 
magnitude $\delta$ to a given, typical path integral history.\footnote{Recall that the configurations of (C)DT in $d$ dimensions are simplicial manifolds made of $d$-simplices (a $0$-simplex is a vertex, a $1$-simplex an edge, a 2-simplex a triangle, a 3-simplex a tetrahedron, etc.) 
with all edges of identical length, up to an overall multiplicative factor distinguishing space- from time-like link lengths.} 
This operation decimates the number of simplicial building blocks and generates a
coarse-grained triangulation with edge length $\delta \geq 2$ in lattice units (i.e.\ setting $a= 1$), for a range of $\delta$.

During the coarse-graining step we monitor for effective topology changes affecting the number of connected components, loops or higher-dimensional voids. 
For example, the original geometry may contain \textit{necks}, which by definition are closed loops of edge length $\leq\delta$, whose length is minimal with respect
to local deformations\footnote{more precisely, local edge reroutings, since any path has to run along discrete lattice edges}.  
After the coarse-graining, these necks will typically be very thin or even pinch to a point, such that
the two pieces of geometry on either side of the neck or pinching point can be regarded as effectively disconnected, corresponding to an effective topology change.
As another example, consider a Swiss cheese-style quantum geometry. Depending on how the coarse-graining is set up, holes of
linear size $\leq\delta$ may disappear in an effective sense at coarse-graining scale $\delta$.

\section{Quantum geometry and homology}
\label{sec:hom}

An attractive feature of this construction is the availability of powerful open-source software to compute the homology of large simplicial complexes, which includes 
the triangulations that are generated by our coarse-graining procedure. The larger context for these mathematical and numerical techniques 
is the field of topological data analysis (TDA), whose main objective is to characterize and analyze large sets of data in terms of a basic set of topological and geometrical properties that can be associated with them (see \cite{Carlsson2009,Munch2017} for motivation and introduction).

In a typical application of TDA, the input is a point cloud, i.e.\ a finite set of data points in a linear space $\mathbb{R}^n$, which may be of very high dimension, where each pair of data points has a mutual distance induced by the Euclidean metric on $\mathbb{R}^n$. 
The point cloud together with this distance matrix is then converted to a triangulated geometric object, more precisely, a simplicial complex \cite{Hatcher}. 
The latter is defined as a set $\cal K$ of simplices such that any subsimplex (``face") of an element in $\cal K$ is again in $\cal K$, and the intersection of two
elements $\sigma_1$, $\sigma_2$ of $\cal K$ is either empty or a subsimplex of both $\sigma_1$ and $\sigma_2$.
There are several ways of obtaining such a complex, but a standard strategy is to draw balls of radius $\epsilon$ around the points in the cloud.
Whenever $d+1$ of these $\epsilon$-balls have a nonempty intersection, the $d$-simplex spanned by their centre points is added to the simplicial complex, forming a so-called Čech complex.
In \textit{persistent homology}, an important methodology in TDA, one studies the homology of the simplicial complexes associated with a 
given point cloud 
as a function of the spatial resolution $\epsilon$, and is particularly interested in the topological aspects that are stable as $\epsilon$ is varied 
(see \cite{roadmap} for a concise description).

We borrow some of these ideas and adapt and apply them to the quantum geometries generated in lattice quantum gravity. In the present work,
we treat two toy models of quantum gravity in 2D explicitly, providing a proof of principle for the methodology. 
As far as we are aware, TDA tools have not been applied previously in quantum gravity, with the exception of \cite{Obster2018}, where they are employed
in search of a geometrical spacetime interpretation of certain tensor decomposition data in the context of a canonical tensor model. 
In terms of ingredients, the present work also has some similarities with \cite{Major2009}, 
which aimed to recover aspects of the homology of continuum spacetime from fundamentally discrete causal sets, invoking simplicial complexes at an intermediate stage of the computation.\footnote{We thank R.\ Sorkin for bringing this reference to our attention.}  
In adjacent fields, persistent homology
has been applied in string theory to study string compactification spaces and flux vacua (see \cite{Ruehle2020} for a review), and in classical cosmology to
describe the large-scale matter distribution of the universe (see e.g.\ \cite{Wilding2021} and references therein). 

We will compute the expectation values of Betti numbers of coarse-grained quantum geo\-metries as a function of the coarse-graining scale $\delta$, which
plays the role of the resolution $\epsilon$ mentioned above. The Betti numbers are integers $\beta_k$ counting the number of $k$-dimensional holes of a 
simplicial complex $\cal K$ or, more formally, measuring the rank of its $k$th homology group $H_k({\cal K})$ \cite{Edelsbrunner2010}.

Our starting point are the triangulated configurations of the gravitational path integral on the lattice, given by what in the literature are variably
called piecewise flat, simplicial or combinatorial $d$-manifolds \cite{ACMbook,review1,Gallier},
which in addition to their topological properties also carry metric properties, by virtue of length
assignments to the edges of their constituent $d$-simplices.\footnote{Note that for the two-dimensional Euclidean dynamical triangulations considered 
in Sec.\ \ref{sec:eucl} below we used
a slight generalization of combinatorial manifolds, the \textit{restricted degenerate ensemble} in the classification of \cite{Loll2024}, 
where a pair of vertices can be connected by more than one edge. \label{fn1}} By assumption, the interior of each $d$-simplex is flat, which entails that
its metric properties are uniquely determined by its edge lengths. A simplicial manifold
satisfies the condition that the \textit{link} of each of its $i$-simplices, $i\in [0,1,\dots,d-1]$ is homeomorphic to the sphere $S^{d-i-1}$.
\footnote{See \cite{Gallier} for a definition of ``link" (not to be confused with link in the sense of edge) and further technical details, and \cite{Loll2024} for a discussion and illustration of this and other, less stringent, regularity conditions on triangulations that have been used in 2D quantum gravity.}  
This is the simplicial analogue of the manifold condition that the neighbourhood of each point is homeomorphic to an open subset of 
$\mathbb{R}^d$.
From the point of view of the underlying simplicial complex, it implies that every $i$-simplex is defined \textit{uniquely} by a set of $i+1$ \textit{distinct} 
vertices, which is also an input requirement for the ``simplex tree" simplicial complex module of the GUDHI library \cite{gudhi} we used to determine 
the Betti numbers of the coarse-grained triangulations.

\section{Coarse-graining triangulations}
\label{sec:coarse}

For a given triangulated configuration $T$ of the gravitational path integral, our coarse-graining procedure at resolution $\delta$ consists 
of three steps: (i) select a subset ${\cal S}_\delta$ of the set $V(T)$ of all vertices of $T$, such that the link distance between typical nearest neighbours -- the number of edges in the shortest path connecting the two vertices -- is of the order $\delta$, (ii) construct the Voronoi cells and Voronoi decomposition of ${\cal S}_\delta$,
and (iii) construct the dual of the Voronoi decomposition, which is the searched-for coarse-grained Delaunay triangulation, whose Betti numbers are subsequently being determined.
To illustrate the procedure and some of the subtleties involved, we will next describe these steps in detail for two-dimensional equilateral
triangulations, which feature in the lattice formulations of both Euclidean quantum gravity and causal, Lorentzian quantum gravity after performing
the Wick rotation.
Elements of our construction below are similar to a coarse-graining proposed earlier in \cite{Henson2009}, albeit with a different motivation and application in mind.

\subsection{Creating a vertex sample}
\label{sec:vertex}

Our aim is to construct a subset ${\cal S}_\delta$ of vertices which samples the vertex set $V(T)$ of the 
triangulation $T$ evenly at a given scale $\delta\in\mathbb{N}$.
To achieve this, we use a construction loosely analogous to that of Poisson disk sampling on smooth manifolds \cite{Wang2021, Dunbar2006}. 
Defining an ``open'' geodesic ball of radius $\delta$ centred at a vertex $v$ by $B_\delta (v)\! :=\! \{ v' \in V(T)| d(v, v') < \delta \}$, where $d$ denotes 
the (integer-valued) link distance, we will require that
(i) $\bigcup_{v \in \cal{S}} B_\delta(v)\! =\! V(T)$, i.e.\ each vertex of $T$ lies in at least one $\delta$-ball, and
(ii) for any pair $v, v'\! \in\! {\cal S}_\delta$ with $v\! \not=\! v'$, $v' \! \notin\! B_\delta (v)$, i.e.\ no vertex from the sample ${\cal S}_\delta$ lies in the
    $\delta$-ball of another sample point, which also implies that $d(v,v')\geq\delta$.
In addition, we define the $\delta$-annulus $A_\delta (v)$ of a vertex $v$ as the set $A_\delta (v)\! :=\! \{ v' \in V(T)| \delta\leq d(v, v') < 2\delta \}$.

To generate an evenly distributed 
vertex sample, we work with three dynamical sets\footnote{By construction, the algorithm does not add vertices that were already present in $\cal{S}_\delta$ and ${\cal{S}}_{new}$; these sets may be implemented computationally as dynamical arrays. The set $V_{cov}$ can be implemented as a Boolean array.}: the set ${\cal S}_\delta$, which at the end of the algorithm will be the 
searched-for sample set, the set ${\cal S}_{new}\subset {\cal S}_\delta$, 
which consists of sampled vertices whose annulus still needs to be explored, and the set $V_{cov}\subset V(T)$ of vertices that
have already been covered by geodesic balls.

At the outset, these lists are empty. After picking an initial vertex $v_0\in V(T)$, 
the algorithm proceeds as follows:
\begin{itemize}
    \item[1.] Add $v_0$ to both ${\cal S}_\delta$ and ${\cal S}_{new}$ as a newly sampled vertex and add all vertices contained in the ball 
    $B_\delta (v_0)$ around $v_0$ to $V_{cov}$, implying that those vertices have been covered.
    \item[2.]  Remove a randomly\footnote{Here and elsewhere, ``random" means according to a uniform distribution over all elements in the set.} chosen vertex $v$ from ${\cal S}_{new}$ and determine its annulus $A_\delta (v)$. Then
    \begin{itemize}
    \item[A.] select a random vertex $u$ from $A_\delta (v)\cap (V \backslash V_{cov})$, i.e.\ the part of the annulus that is not yet covered by
    geodesic balls;
    \item[B.] add $u$ to both ${\cal S}_\delta$ and ${\cal S}_{new}$ and add all vertices of the $\delta$-ball around $u$ to $V_{cov}$;
    \item[C.] if $A_\delta (v)\cap (V \backslash V_{cov})$ is nonempty, repeat from 2a.
    \end{itemize}
    \item[3.] If ${\cal S}_{new}$ is nonempty, repeat from 2.
\end{itemize}

\begin{figure}[ht]
    \centering 
    \includegraphics[width=\linewidth]{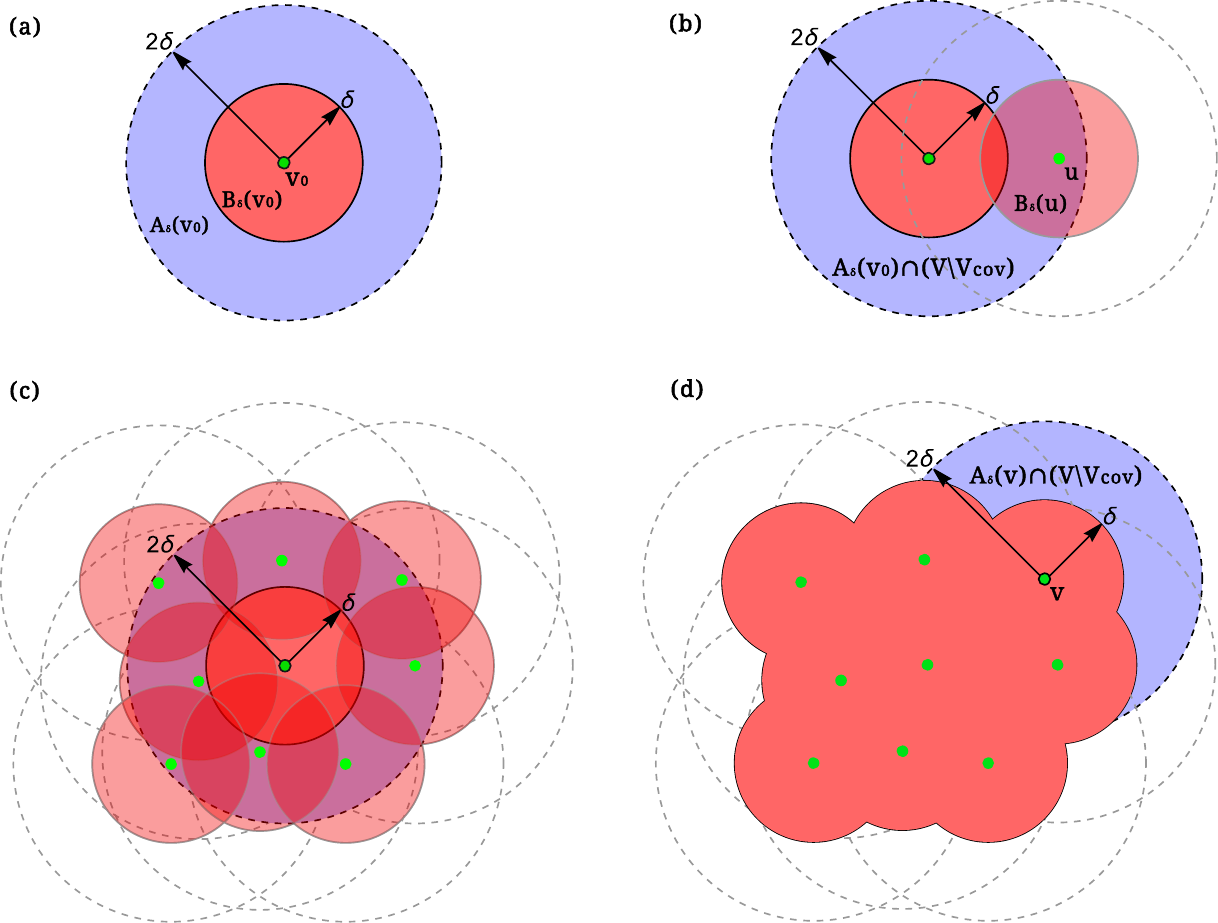}
\caption{
Illustrating the covering algorithm: (a) initial vertex $v_0$, together with its $\delta$-ball $B_\delta (v_0)$ (red) and $\delta$-annulus $A_\delta (v_0)$ (blue); 
(b) random vertex $u$ from the part of the annulus not yet covered, together with its $\delta$-ball $B_\delta (u)$ (pink);
(c) after several iterations, all vertices in $A_\delta (v_0)$ have been covered; (d) algorithm continues with a random vertex
$v\in {\cal S}_{new}$ whose annulus is not yet covered. 
}
\label{fig:well-spread-sample-algorithm}
\end{figure}

Fig.\ \ref{fig:well-spread-sample-algorithm} is a schematic, planar illustration of elements of this algorithm. 
Note that we only indicate the vertices that end up in the sample ${\cal S}_\delta$ (in green), and not any of the other vertices contained in the geodesic 
$\delta$-balls around them. In Fig.\ \ref{fig:well-spread-sample-algorithm}d, the combined red region represents the set $V_{cov}$ up to this stage, while the blue
region contains the vertices that still need to be covered by the algorithm.

At the end of this process, we have created a vertex sample ${\cal S}_\delta$ which by construction satisfies condition (ii) from the beginning of Sec.\ \ref{sec:vertex}. 
To see that it also satisfies (i), we can argue by contradiction. Assume that (i) does not hold and 
there therefore exists a vertex $w\notin V_{cov}$. This vertex must have a distance of at least $2\delta$ from all elements of ${\cal S}_\delta$,
since the algorithm by construction covers all vertices inside this radius. 
Since $T$ by assumption is connected, there is a (possibly nonunique) vertex $p\in{\cal S}_\delta$ which has a finite, minimal link distance
$d_{min}=d(p,w)\geq 2 \delta$ to $w$. In other words, there is a (possibly nonunique) 
chain of $d_{min}$ consecutive edges between $p$ and $w$ that form a path of this
minimal length. At $2\delta -1$ steps from $p$ along this chain there is therefore a vertex $q$, which lies in the annulus $A_\delta(p)$, and
has distance $d_{min}-2\delta +1 $ from $w$. Although $q$ itself may not be in ${\cal S}_\delta$, it must lie in a 
$\delta$-ball of some other vertex $q' \in {\cal S}_\delta\cap A_\delta(p)$, since by construction the annulus $A_\delta(p)$ 
is completely covered by $\delta$-balls. From $d(q,q')\leq \delta-1$ it then follows that $d(q',w)\leq d_{min}-\delta$, which implies that
the sample vertex $q'\in {\cal S}_\delta$ is closer to $w$ than $p\in {\cal S}_\delta$, contradicting our original choice of $p$. Thus we
have shown that condition (i) holds too.  
To summarize, we have shown that our construction delivers an evenly spread set of sample points.

\subsection{Constructing Voronoi cells}
\label{sec:voronoi}

Given a sample ${\cal S}_\delta$ for a given triangulation $T$, our next step is to partition the set $V(T)$ of all vertices into Voronoi cells,
and subsequently extend this to a Voronoi decomposition of the entire two-dimensional triangulation.  
The Voronoi construction is most familiar as a prescription for decomposing (``tessellating") the Euclidean plane into $n$ cells 
associated with a set of seed points
$\{ p_i ,\, i=1,2,\dots n \}$, where the Voronoi cell associated with a given point $p_i$ is given by the set of all points closer to $p_i$ than to 
any other point $p_j$, $j\not= i$, or at an equal distance to such points. An analogous prescription can be used on a graph
of the type considered here (see, e.g.\ \cite{Erwig2000}): the cell associated with a given vertex $v\in {\cal S}_\delta$ 
will by definition consist of all vertices closer to $v$ (in terms of the link distance) 
than to any other $v' \in {\cal S}_\delta$. If a vertex has the same distance to more than one
seed vertex, a random assignment is made, as specified below. 

\begin{figure}[t]
\newcommand{\triangleplot}[2]{
    \begin{minipage}{0.2\linewidth}
    \begin{minipage}{\linewidth}
    \includegraphics[width=\linewidth]{#2}
    \end{minipage}\\[2mm]
    \begin{minipage}{\linewidth}
    \centering
    \Large #1
    \end{minipage}
    \end{minipage}
}
\centering
\triangleplot{type 0}{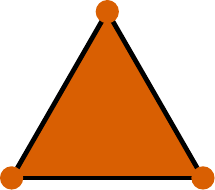}
\hspace{5mm}
\triangleplot{type 1}{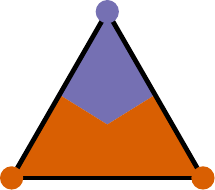}
\hspace{5mm}
\triangleplot{type 2}{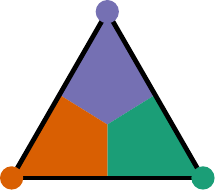}
\caption{Triangles can have three types of colouring, depending on the colouring of their vertices, generated during the decomposition
of vertices into Voronoi cells.}
\label{fig:threetri}
\end{figure}

\begin{figure}[h!]
\centering
\includegraphics[width=0.95\textwidth]{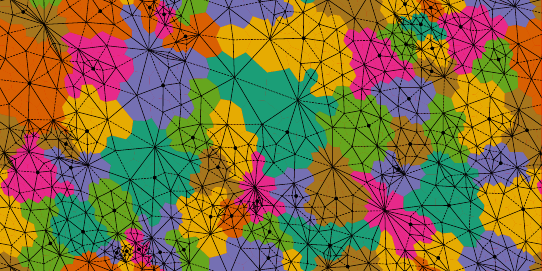}
\caption{Voronoi decomposition into cells of a 2D CDT configuration on a torus, for $\delta =3$,
using a Tutte or barycentric embedding, where each vertex is located at the barycentre of its neighbours.
Dashed and solid lines represent space- and
timelike edges respectively, and each black dot is a seed vertex for a cell of a given colour. 
Note the toroidal periodicity for opposite sides of the rectangle.}
\label{fig:voro}
\end{figure}

Our seed points will be the vertices in ${\cal S}_\delta$, and the ``cell" corresponding to a given seed vertex $v$ is found by performing 
a breadth-first search of its neighbouring vertices for increasing link distance $d=1,2,\dots$ from $v$. 
Algorithmically, we perform the breadth-first search radially outward from all seeds simultaneously, in unit steps. 
This means that at the $d$th step we assign vertices to the cell of the nearest seed and
make a random assignment (with uniform probability) in case there is more than one seed at distance $d$, before moving on to 
step $d+1$. This guarantees that the Voronoi cells defined on the entire triangulation we will construct in the next step are connected.

We now extend the colouring to the triangles of $T$ by dividing each triangle into three ``dual" area segments of the same colour
as the associated corner vertex. Three types of triangle colouring are possible, depending on whether all three vertices have the same
colour (type 0), exactly two vertices have the same colour (type 1), or all three have different colours (type 2), see Fig.\ \ref{fig:threetri}.
For type 1, the triangle contains a boundary between two (two-dimensional) Voronoi cells of different colour, while for type 2
it contains a point where three cells meet, as well as three pairwise boundaries. Fig.\ \ref{fig:voro} shows a typical decomposition 
of a two-dimensional CDT configuration
into such Voronoi cells, where we have associated one cell to each vertex of the evenly distributed sample set ${\cal S}_\delta$.
In this example, each cell has the topology of a disc.  

\newpage
\subsubsection{Properties of Voronoi cells}
\label{sec:cells}

As illustrated by Figs.\ \ref{fig:threetri} and \ref{fig:voro}, in our construction at most three Voronoi cells can meet in a point. Such a triple point is always located 
at the centre of a triangle, which can also be viewed as a vertex of the trivalent graph dual to the triangulation $T$. Likewise,
the boundaries between neighbouring cells consist of edges of this dual graph. A sequence of dual boundary edges between two 
adjacent (along the boundary) triple points we will call a \textit{boundary segment}.

The topology of a typical cell is that of a disc, but any connected subset of the original
triangulated manifold can in principle occur. Since we will never consider a cell of maximal volume that covers the entire triangulation,
a cell will always have one or more boundaries, each of which is topologically a circle. 
An important case that appears frequently when coarse-graining 2D Euclidean DT configurations is that of an annulus, i.e.\ a disc with a hole. 
This can happen when a cell wraps completely around a ``thin neck" of the triangulation, as illustrated by Fig.\ \ref{fig:wrap}. As we will see in
Sec.\ \ref{sec:propdel} below, this is associated with a ``pinching" of the coarse-grained triangulation.

\begin{figure}[t]
\centering
\includegraphics[clip, trim=0mm 14mm 0mm 14mm, width=0.5\textwidth]{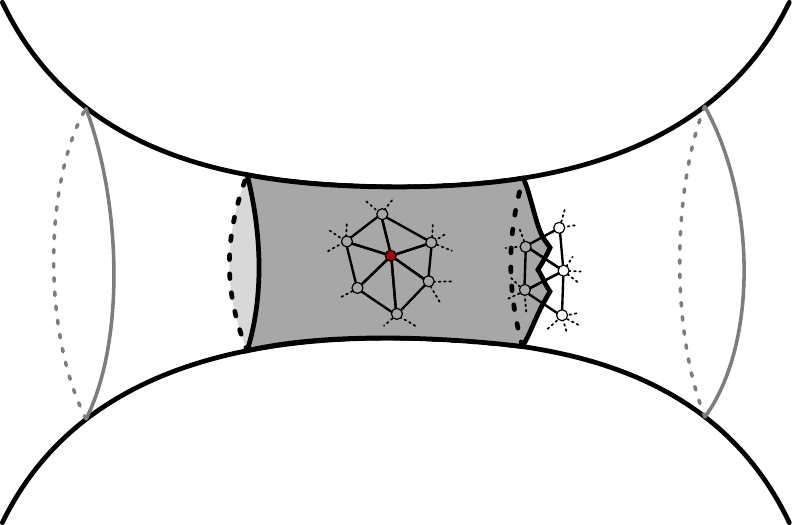}
\vspace{3mm} 
\caption{Voronoi cell associated with a seed vertex (in red), with the topology of an annulus (shaded region), 
wrapping around a thin neck of the triangulation and shown
in a three-dimensional embedding.
}
\label{fig:wrap}
\end{figure}

\begin{figure}[ht]
\centering
\includegraphics[width=0.4\textwidth, angle=90]{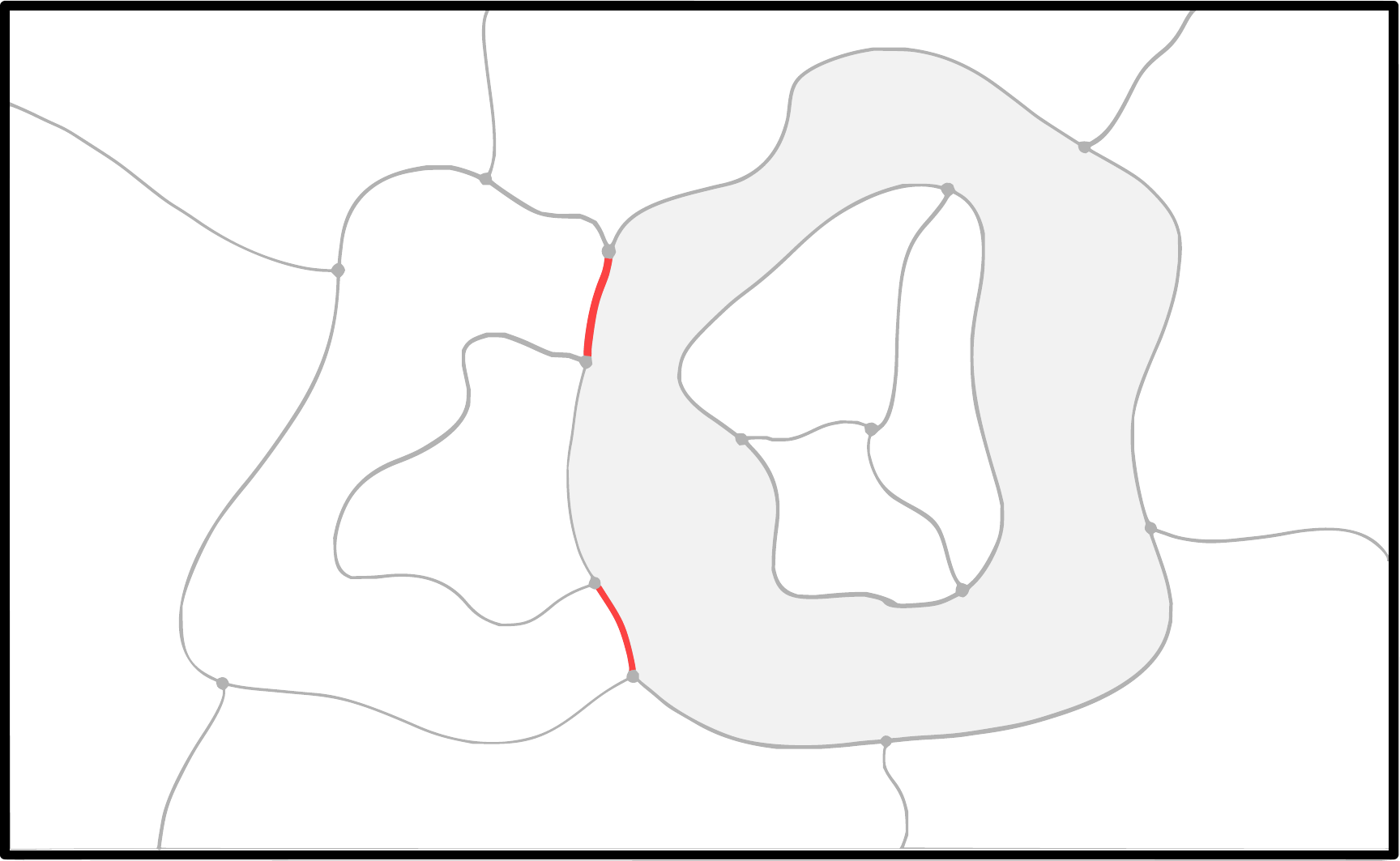}
\caption{An annulus-shaped (shaded gray) and a disc-shaped 
Voronoi cell meet along two disconnected boundary segments (in red). This schematic, planar drawing only shows boundary segments and triple points.
}
\label{fig:wrap2}
\end{figure}

Note also that two cells can meet along several, disconnected boundary segments, as illustrated by Fig.\ \ref{fig:wrap2}. 
This situation can occur when two cells wrap around a thin neck. It can also happen that an entire circular boundary
consists of a single boundary segment only, without any triple points, for example, when a disc-shaped cell is surrounded by an annulus. 

\subsection{Constructing the dual Delaunay triangulation}
\label{sec:delaunay}

Our next objective is to obtain a coarse-grained triangulation $T_\delta$, which is dual to the Voronoi decomposition described in Sec.\ \ref{sec:voronoi}, in a sense we will describe below. We are primarily interested in the effective topology of $T_\delta$, 
where features of up to a linear size $\delta$ are ``disregarded". How precisely this is implemented is ultimately a matter of choice. 
Our general strategy will be to coarse-grain in a way that is maximally topology-preserving, and such that any localized topology 
changes that occur in this process are well controlled.\footnote{Note in particular that for the choice $\delta=1$ (no coarse-graining) the procedure of Secs.\ \ref{sec:vertex}--\ref{sec:delaunay} leads to a dual Delaunay triangulation $T_1$ that reproduces the original triangulation $T$.} This requires a good understanding of the geometric and topological properties of a local 
neighbourhood of the triangulation, which is rather straightforward in dimension two, but more involved in higher dimensions. 

The manner in which we will set up the coarse-graining leads to a slightly generalized notion of ``triangulation" for $T_\delta$, compared to
the one we started with for $T$. We will nevertheless call it a \textit{Delaunay triangulation}\footnote{also sometimes called a Delaunay complex}, 
because it is inspired by a standard construction of the same name in the Euclidean plane (or, more generally, $\mathbb{R}^n$),
with the following
properties understood. Firstly, $T_\delta$ will in general no longer be
a topological manifold, in the sense that it will not look two-dimensional in every point. This happens when 
thin necks of the original triangulation pinch to a point or collapse to a sequence of one or more edges, which are not part of any triangle.
In turn, these edge sequences can also branch into tree-like structures (see also Fig.\ \ref{fig:bubbles} below). 
Secondly, even away from points which do not have a
$\mathbb{R}^2$-like neighbourhood, the two-dimensional triangulation will in general not be a simplicial manifold -- according to our definition
in Sec.\ \ref{sec:hom} above -- because it can happen that two edges share the same endpoints, forming a closed loop of length 2. An explicit example will be presented in Sec.\ \ref{sec:propdel} below. 

The coarse-grained Delaunay triangulation $T_\delta$ inherits the connectivity properties of its constituent vertices, edges and triangles from 
the Voronoi decomposition, and is subsequently endowed with metric properties
by assigning a uniform length $\delta$ to all of its edges and by declaring its edges as straight and its triangles as flat. 
However, unlike reference \cite{Henson2009}, we will not be
interested in the nontrivial curvature properties of $T_\delta$, but only in its effective topology.
The first step to obtaining the Delaunay triangulation identifies subsets of the Voronoi decomposition as dual vertices, edges and
triangles, keeping track of their neighbourhood relations:  
\begin{itemize}
\item[(i)] with each Voronoi cell we associate a ``vertex" inside the cell. This vertex, dual to the Voronoi decomposition, 
can be thought of as midpoint of the cell, but its actual position is immaterial, since in this step we only keep track of the connectivity of the
dual simplicial building blocks of what will become the Delaunay triangulation.
\item[(ii)] With each boundary segment shared by a pair of Voronoi cells we associate an ``edge", whose two endpoints are
dual vertices as described in (i). We can construct this dual edge as a simple path that connects the two dual vertices and traverses the boundary segment at some (arbitrarily defined) midpoint. The details of this choice are again unimportant.
\item[(iii)] With each vertex of the Voronoi decomposition we associate a ``triangle", whose edges are the duals to the three boundary segments
meeting at the vertex, as described in (ii). 
\end{itemize}
Note that as subsets of the original triangulation $T$, the ``edges" identified in (ii) are in general not straight, and the ``triangles" 
in (iii) are not flat. The searched-for coarse-grained Delaunay triangulation $T_\delta$ is now \textit{defined} as the piecewise flat space
consisting of the abstract simplicial building blocks identified in steps (i)--(iii), together with their connectivities, where in addition we 
assign metric properties, namely that all edges are straight and of length $\delta$ and all triangles carry a
flat Euclidean metric structure induced by their three boundary edges. 

\begin{figure}[t]
\centering
\includegraphics[width=0.4\textwidth, angle =90 ]{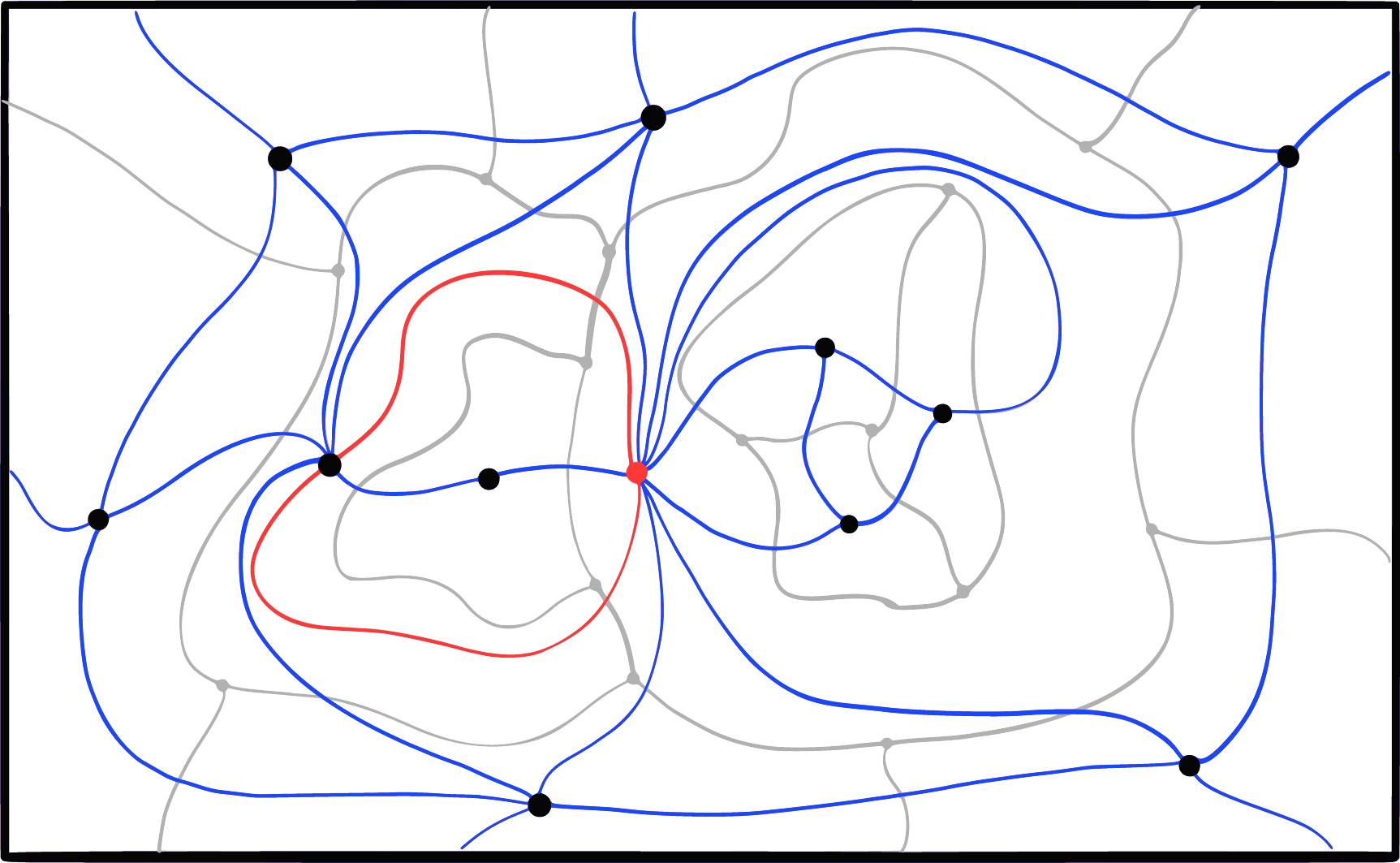}
\caption{A Voronoi decomposition (black) and its dual Delaunay triangulation (grey). Highlighted in red are a pinching vertex and a closed loop
consisting of two edges, dual to the annulus-shaped Voronoi cell and the red boundary segments of Fig.\ \ref{fig:wrap2} above. 
}
\label{fig:dualdel}
\end{figure}

This construction is illustrated by Fig.\ \ref{fig:dualdel}, which shows a piece of a Voronoi decomposition (the same as in Fig.\ \ref{fig:wrap2} above) 
and its dual Delaunay triangulation in
a schematic, planar representation. The dual vertices at the centre of Voronoi cells and the dual edges, connecting pairs of dual
vertices, are drawn in grey. For better readability, we have not coloured the triangles of the Delaunay triangulation, 
which are dual to trivalent vertices of the Voronoi decomposition and whose boundaries are given by closed loops of three grey edges each.

\subsubsection{Properties of the Delaunay triangulation}
\label{sec:propdel}

Fig.\ \ref{fig:dualdel} also illustrates some of the irregular features of the Delaunay triangulation $T_\delta$ 
we already mentioned above, and which would not occur
for a Voronoi decomposition of the Euclidean plane. Firstly, there are the two cells, one with the topology of an annulus, the other
one with the topology of a disc, which share two disconnected boundary segments. According to prescription (ii), this leads to
two distinct dual edges between the dual vertices associated with these two cells. They form a closed loop consisting of
two edges in the Delaunay triangulation, thereby violating the simplicial manifold conditions. 

Secondly, the vertex marked in red in Fig.\ \ref{fig:dualdel}, which is dual to the cell with the topology of an annulus, is what we call a 
pinching vertex. As was already mentioned in 
Sec.\ \ref{sec:eff}, it is associated with the presence of a neck in the original triangulation $T$, which in the coarse-grained Delaunay triangulation 
gets pinched down to a point. As a consequence, its neighbourhood is
no longer homeomorphic to an open subset of $\mathbb{R}^2$, thereby violating the manifold condition.

For our purposes, the most important feature of the coarse-graining process is the pinching that occurs at the location of an annulus.
Since the annulus, viewed as a cell of the Voronoi decomposition, by definition does not have any boundary segments in its interior, 
the dual Delaunay triangulation does not contain any edges that 
lie entirely
inside the annulus.
This implies that the coarse-grained triangulation is no longer a two-dimensional manifold at the pinching vertex.\footnote{For added clarity,
one can envisage the pinching as a two-step process: (i) by hand, introduce a single dual edge that links the dual vertex inside the annulus to itself and winds around the
annulus once; this preserves the manifold character of the resulting dual triangulation. (ii) Shrink the dual loop of length 1 to a point.}
Since the annulus separates the Voronoi cells into those lying in- and outside the annulus\footnote{This is not necessarily true when $T$ has
a global topology with noncontractible loops, but will nevertheless lead to a local pinching.}, the resulting Delaunay triangulation will
in general have two parts located on either side of the pinching vertex, which are connected to each other only at the vertex.

\begin{figure}[t]
    \centering
    \includegraphics[width=0.4\linewidth]{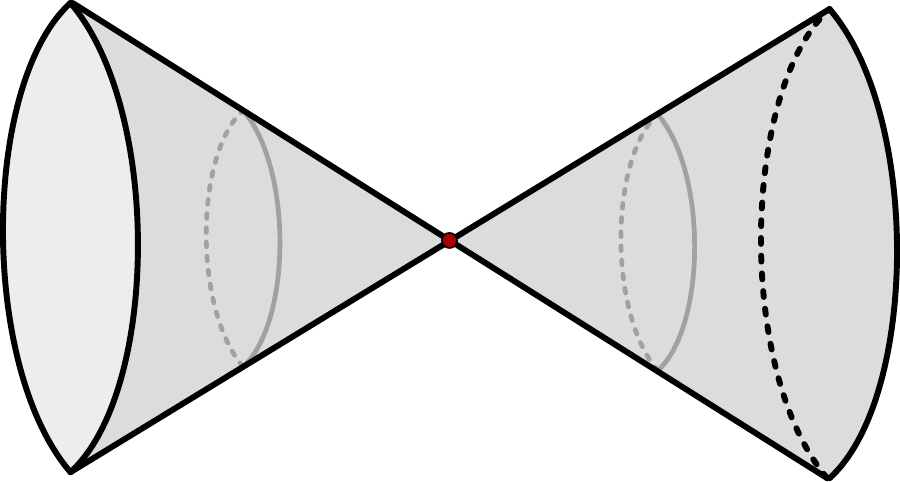}
    \vspace{3mm} 
    \caption{Schematic depiction of a piece of coarse-grained Delaunay triangulation associated with a pinching vertex (red) at which two cones meet. It is dual to the Voronoi cell
    of Fig.\ \ref{fig:wrap}, whose seed vertex can be identified with the pinching vertex here. }
    \label{fig:double-cone}
\end{figure}

\begin{figure}[t]
\vspace{3mm}
\centering
\includegraphics[width=.8\linewidth]{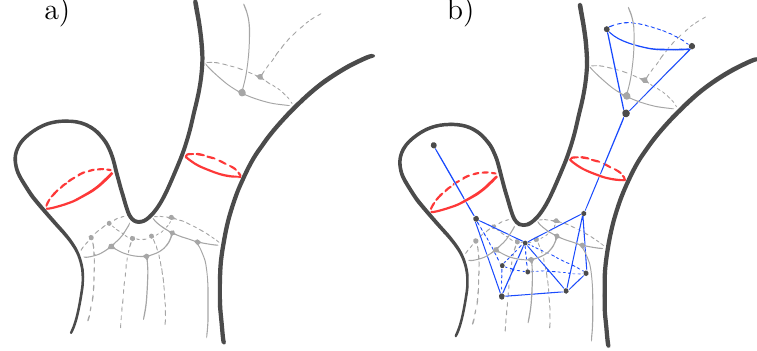}
\caption{(a): Schematic depiction of a Voronoi decomposition, containing two boundary segments without triple points marked in red. 
(b): ditto, with the Delaunay triangulation added (blue), containing two loose edges dual to the marked boundary segments.
}
\label{fig:loose3d2d}
\end{figure}

In the simplest case, the neighbourhood of such a pinching point is homeo\-morphic to that of two two-dimensional cones meeting at their tips (Fig.\ \ref{fig:double-cone}).
When the Voronoi cell has the topology of a disc with $h>1$ holes instead of $h=1$ (the annulus), the situation generalizes in the simplest case to that of a 
meeting point of the tips of $h+1$ cones.
Another type of generalization -- beyond these simplest cases -- occurs when the Voronoi decomposition contains two or more nested annuli. 
The shared boundary between two such annuli consists of a single boundary segment without any trivalent vertices located on it, 
which implies that the edge dual to this
segment is not part of any triangle in $T_\delta$, but is instead what we will call a ``loose" edge. 
The dual Delaunay triangulation has then not just a pinching vertex, but an entire ``pinching edge", along which the triangulation is locally one-dimensional. 
This is illustrated by Fig.\ \ref{fig:loose3d2d}a, which shows two instances of boundary segments without vertices (highlighted in
red) of the Voronoi decomposition. The corresponding dual Delaunay triangulation is added in blue in Fig.\ \ref{fig:loose3d2d}b.
The boundary segment between a disc-shaped cap and an annulus on the left branch of the geometry leads in the Delaunay 
triangulation to a loose 
edge ending in a single vertex, and the boundary segment between two annuli on the right branch leads to a loose edge 
connecting two different parts of the coarse-grained Delaunay triangulation.

\begin{figure}[t]
\centering
\includegraphics[width=0.6\linewidth]{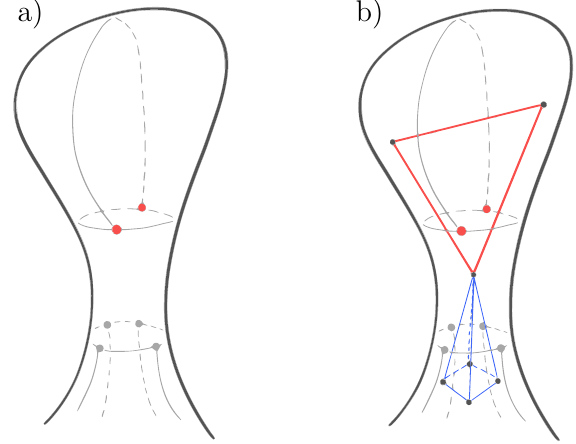}
\caption{(a): Voronoi decomposition where the same three Voronoi cells are shared by two distinct triple points (red). (b): In the dual Delaunay triangulation, this results in a pillow, consisting of two triangles (red) sharing the same edges and vertices.}
\label{fig:pillow-diagram}
\end{figure}

Yet another irregular feature one can encounter during coarse-graining are so-called ``pillows". By definition, a
pillow -- in this case of a Delaunay triangulation -- consists of two distinct triangles, which share the same three
(distinct) vertices and three (distinct) edges. 
This violates the combinatorial character of a simplicial manifold, since the triangles are not
uniquely characterized in terms of their vertices. An example of how this can occur is illustrated by Fig.\ \ref{fig:pillow-diagram}. 
It shows a Voronoi decomposition with the particular property that there are two distinct triple points (marked in red) at which the same three Voronoi cells meet; in the case at hand, these are two discs and one annulus (Fig.\ \ref{fig:pillow-diagram}a). By virtue of our algorithm, this leads to two distinct Delaunay triangles forming a pillow, highlighted in red
in Fig.\ \ref{fig:pillow-diagram}b. In the example shown, it is
connected to the remainder of the Delaunay triangulation by a pinching vertex, associated with the annulus of the Voronoi decomposition. The occurrence of this type of local structure is relevant for our analysis of the effective topology below, since a pillow is topologically a two-sphere and therefore contributes 
to the count of the Betti number $\beta_2$.

\begin{figure}[ht]
\vspace{3mm}
\centering
\includegraphics[width=0.7\textwidth]{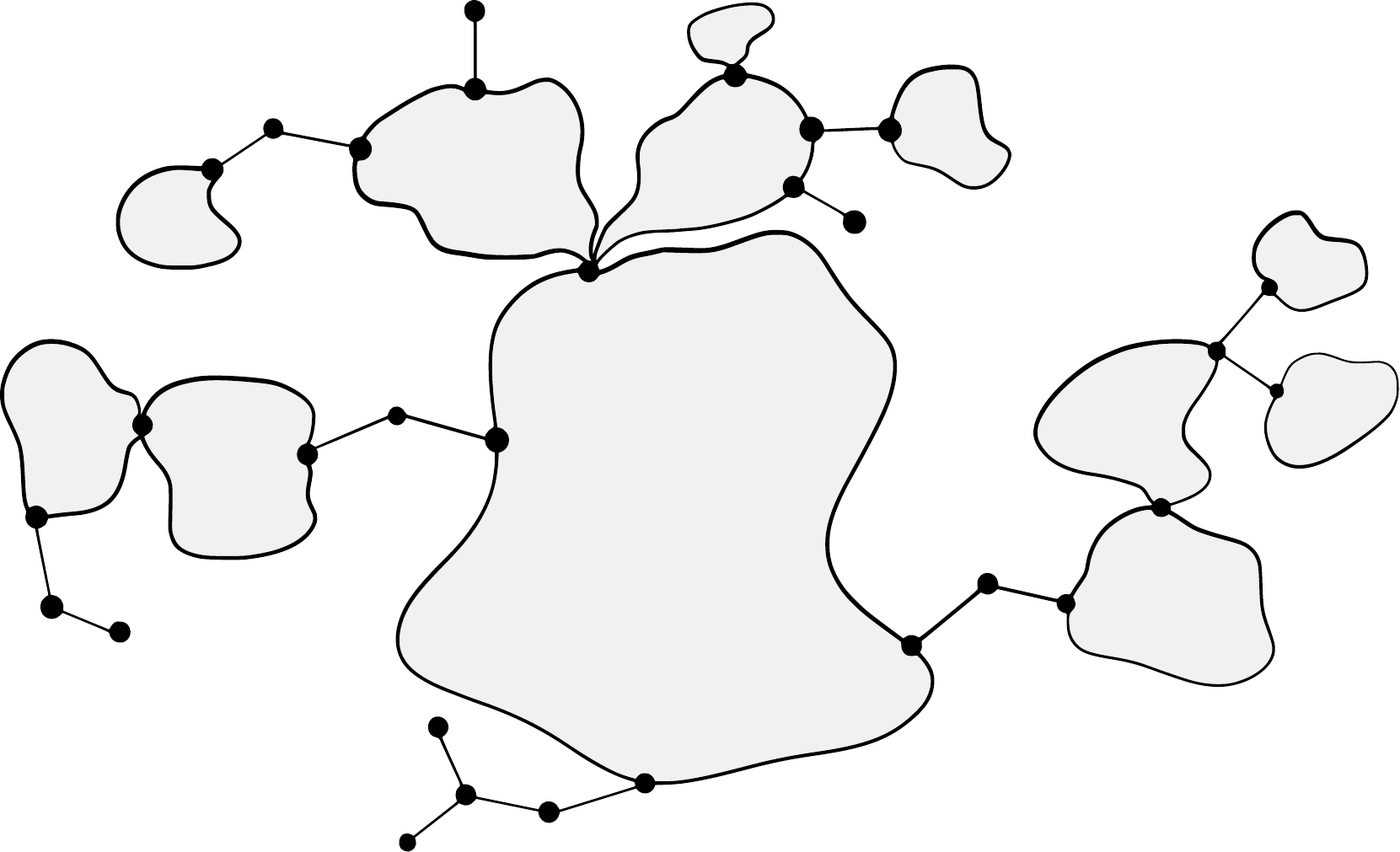}
\vspace{3mm}
\caption{Diagram illustrating the topological features that can occur in a coarse-grained Delaunay triangulation $T_\delta$ obtained from a triangulation $T$ of
spherical topology. Each circle or ``bubble" filled with gray lines represents a spherical triangulation of arbitrary size. Loose edges can connect between such bubbles or be
attached to them as outgrowths.}
\label{fig:bubbles}
\end{figure}

Note that by construction of the Delaunay triangulations $T_\delta$, no one-dimensional boundaries can appear during coarse-graining, 
and it also cannot happen that parts of $T_\delta$ become disconnected.  
Consequently, the overall topology of $T_\delta$ consists of two-dimensional spherical ``bubbles" and generalized bubbles with more complicated shapes\footnote{e.g.\ enclosed by surfaces of higher-genus, depending on the topology of the original triangulations $T$}, connected to each other at pinching vertices or by (sequences of)
loose edges, where multiple bubbles and/or loose edges can meet at a given pinching vertex, as illustrated by Fig.\ \ref{fig:bubbles}. The bubbles of minimal size are exactly the pillows we have just introduced.  

\section{Measuring the effective homology of quantum geometry}
\label{sec:effhom}

In the context of quantum gravity, the effective homology of space(-time) introduced above has the status of a diffeomorphism-invariant observable.
In what follows, we will measure this observable with the help of Monte Carlo simulations in two toy models of quantum gravity\footnote{The implementation for both models is open-source \cite{DuinCode}, with a repository available at \url{https://gitlab.com/dynamical-triangulation/dyntri-rs.}
}, defined
nonperturbatively as continuum limits of gravitational path integrals regularized on dynamical lattices. We will treat the case of Lorentzian quantum gravity,
formulated in terms of Causal Dynamical Triangulations (CDT), in Sec.\ \ref{sec:lor} and that of Euclidean quantum gravity, formulated in terms of 
Euclidean Dynamical Triangulations (EDT), in Sec.\ \ref{sec:eucl}.
Among other things, this will help us to assess their suitability as carrier
spaces of symmetry.

To set the context, let us recall the computation of Betti numbers of
simplicial complexes, which we will use to analyze the Delaunay
triangulations produced by the coarse-graining described in Sec.\
\ref{sec:coarse} (see e.g.\ reference \cite{Edelsbrunner2010}). It turns out to
be a problem in linear algebra that is amenable to computation and scalable to
large triangulations. 

The setting is that of a $d$-dimensional simplicial complex $\cal K$, where by definition $d$ is the maximal dimension of any of its simplices.
A \textit{p-chain} $c$ is defined as the formal sum $c=\sum_i a_i \sigma_i^{(p)}$ of $p$-simplices $\sigma_i^{(p)}\in{\cal K}$, where for
simplicity we use mod-2 coefficients $a_i$ which can take the values $a_i=0,1$. One can also define simplicial homology over larger finite coefficient fields, but
for our present purposes this choice will not make a difference.\footnote{
    In the GUDHI implementation we have used the default field $\mathbb{F}_{11}$ instead of $\mathbb{F}_2$.
}
The $p$-chains form an abelian group $C_p({\cal K})$ under addition. Next, we define a \textit{boundary map} $\partial_p$, 
which maps $p$-chains to $(p-1)$-chains and on basis elements is defined as
\begin{equation}
\partial_p : C_p\rightarrow C_{p-1},\;\;\; \sigma^{(p)}=\{v_0,v_1,\dots,v_p\} \mapsto \sum_{i=0}^p \{v_0,\dots,\hat{v_i},\dots,v_p\},
\label{bmap}
\end{equation}
where we have represented the simplex $\sigma^{(p)}$ by the list of its $(p+1)$ vertices $v_i$, $\sigma^{(p)}=\{v_0,v_1,\dots,v_p\}$,
and the hat denotes the omission of the $i$th vertex. In geometric terms, eq.\ (\ref{bmap}) says that the boundary of a simplex $\sigma^{(p)}$ is the
sum of its $(p-1)$-dimensional faces. 
The boundary map allows us to define two subgroups of the group $C_p({\cal K})$ of $p$-chains, namely (i) the group $Z_p({\cal K})$ 
of \textit{p-cycles} $c$ satisfying $\partial_p c=0$, and (ii) the group $B_p({\cal K})$ of \textit{p-boundaries} $c$ satisfying 
$c=\partial_{p+1}d$ for some $(p+1)$-chain $d$. It then follows from the fundamental lemma of homology, 
\begin{equation}
\partial_p\partial_{p+1}\, d=0,\;\; \forall d\in C_{p+1}({\cal K}),
\label{lemma}
\end{equation}
``the boundary of a boundary is zero", that the $p$-boundaries form a subgroup of the $p$-cycles, i.e.\ $B_p({\cal K})\subset Z_p({\cal K})$.

The $p$th homology group $H_p$ is then defined as the group of $p$-chains modulo the group of $p$-boundaries, 
$H_p({\cal K}):= Z_p({\cal K})/B_p({\cal K})$. Since all of these groups are also vector spaces, this can be expressed equivalently 
as the quotient vector space
\begin{equation}
H_p({\cal K}) = \mathrm{kernel}(\partial_p)/\mathrm{image}(\partial_{p+1}).
\label{quot}
\end{equation}
Its dimension,
\begin{equation}
\beta_p({\cal K}):= \dim H_p({\cal K})=\dim Z_p({\cal K}) -\dim B_p({\cal K}),
\label{dimh}
\end{equation} 
is called the $p$th \textit{Betti number}.
In other words, the Betti number $\beta_p$ counts $p$-cycles that are \textit{not} $p$-boundaries, which
geometrically can be thought of as $p$-dimensional ``holes". In two dimensions, the case at hand, $\beta_0$ counts the number of
components, $\beta_1$ the number of loops (one-dimensional holes), and $\beta_2$ the number of two-dimensional holes.
We also see that determining the Betti numbers of a simplicial complex
involves the computation of the ranks of the linear maps $\partial_p$. 

As already mentioned in Sec.\ \ref{sec:hom}, we have used the open-source C++ library GUDHI to compute the Betti numbers of the coarse-grained
Delaunay triangulations $T_\delta$. To give an idea of its efficiency, it takes on the order of a second to compute the $\beta_i$ for a triangulation of
size 100$k$. However, there is one small step that must still be taken before this library can be used. Implicit in our definition (\ref{bmap}) of the
boundary map was the unique characterization of each $p$-simplex $\sigma^{(p)}$ in terms of its $p+1$ vertex labels. 
This vertex representation is required by GUDHI as an input format, but is in general not satisfied by our Delaunay triangulations, as we have already seen: 
the marked, closed loop consisting of two dual edges depicted in Fig.\ \ref{fig:dualdel} is an explicit example, since both edges share the same vertices.  
This is merely a technical issue, since the homology of this generalized simplicial complex\footnote{Technically, it is an example of a $\Delta$-complex, which has a well-defined simplicial homology, see 
\cite{Hatcher}.} is perfectly well defined. We resolve this by making small local adjustments to the Delaunay triangulation, without changing its homology, 
such that all of its edges and triangles are uniquely defined through their vertices. Details of this procedure can be found in Appendix B.

\subsection{Lorentzian quantum gravity in D=2}
\label{sec:lor}

We begin by evaluating the expectation values of Betti numbers in coarse-grained spacetimes of resolution $\delta$ in two-dimensional Lorentzian quantum gravity.
The path integral is given by the two-dimensional analogues of relations (\ref{pathint}) and (\ref{seh}), where the integral over the two-dimensional Ricci scalar is a topological invariant
and will be dropped from the action. The lattice-regularized version of this path integral is a discrete sum over CDT with a well-defined
causal structure. After the Wick rotation it reads \cite{Ambjorn1998,review1}
\begin{equation}
Z(\lambda)=\sum_{\mathrm{causal}\, T} \frac{1}{C(T)}\, \mathrm{e}^{-\lambda N_2(T)},
\label{pi2dcdt} 
\end{equation}
where $\lambda$ is the bare cosmological constant, $N_2$ the number of triangles, 
and $C(T)$ denotes the order of the automorphism group of the triangulation $T$, which consists of all
maps of $T$ to itself that preserve all of its neighbourhood relations.

\begin{figure}[t]
\centering
\includegraphics[width=0.35\textwidth]{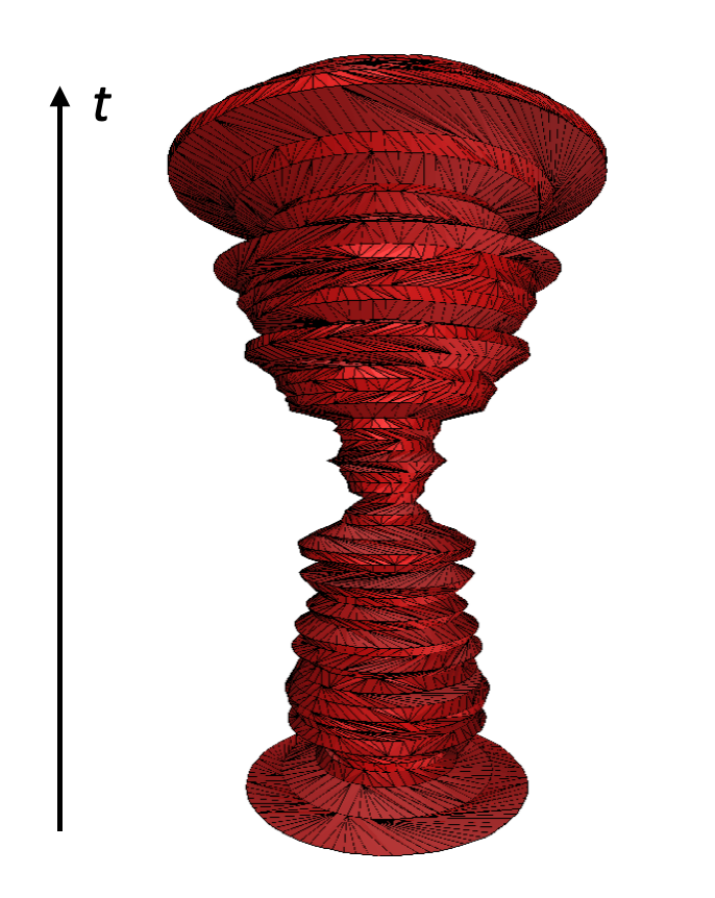}
\caption{Typical CDT configuration, with volume $N_2=4.096$ and time extension $t_\mathrm{tot}=64$, depicting a compact spatial quantum universe whose size fluctuates
in time. In the simulations,
the time direction is cyclically identified.}
\label{fig:cdt-history}
\end{figure}

Recall that each CDT configuration is a sequence of one-dimensional spatial universes labelled by an integer time $t\!\in\! [0,1,2,\dots,t_\mathrm{tot}]$,
where strips assembled from identical, triangular flat building blocks interpolate between adjacent spatial universes of variable size (Fig.\ \ref{fig:cdt-history}).
This stacked structure of the spacetimes is a lattice implementation of global hyperbolicity, a hallmark of the causal structure of CDT that is not present in the histories of EDT.
For convenience we cyclically identify the time direction, $t_\mathrm{tot}\equiv 0$, and use compact spatial universes of $S^1$-topology, such that
all configurations $T$ have the topology of a two-torus $T^2 \equiv S^1\times S^1$. 
The geometric variable characterizing a spatial slice at integer $t$ is its one-dimensional volume, which in lattice units is given by the number $\ell (t)$
of its edges, where we demand $\ell(t)\!\geq\! 3$ to ensure that $T$ is a simplicial manifold. 

As is customary, our simulations have been performed for ensembles of fixed total volume $N_2$, in the present case
for eight equally spaced values $N_2 \in [50k,400k]$. 
For given $N_2$, we have performed measurements for various fixed time extensions $t_\mathrm{tot}$.
This changes the average spatial volume $\bar{\ell}\! =\! N_2/(2 t_\mathrm{tot})$,
which can be interesting to exhibit the dependence of the Betti numbers on global properties of the underlying geometry. 
Following the considerations of \cite{correl}, 
we have used 18 different ratios $r := t_\mathrm{tot}^2/N_2$ for each $N_2$, evenly spaced in the range
$r \in[0.08,0.25]$.
Whenever we study the volume-dependence of a given quantity, we will for simplicity focus on configurations with the intermediate value $r=0.16$.

The expectation values of geometric observables $\cal O$ in this ensemble are given by\footnote{In the remainder of the paper, we will drop the subscript $N_2$
for notational convenience, but it should be understood that all expectation values refer to constant-volume ensembles.}
\begin{equation}
\langle {\cal O}\rangle_{N_2} =\frac{1}{\tilde{Z}(N_2)}   \sum_{\mathrm{causal}\, T|_{N_2}} \frac{1}{C(T)}\, {\cal O}(T),\;\;\;\;\;\; \tilde{Z}(N_2)=  \sum_{\mathrm{causal}\, T|_{N_2}} \frac{1}{C(T)},
\label{expfix}
\end{equation}
where the sums are over CDT configurations of fixed volume $N_2$ and time extension $t_\mathrm{tot}$.
The fixed-volume path integral $\tilde{Z}(N_2)$ is related to the path integral (\ref{pi2dcdt}) for fixed cosmological constant
by a Laplace transform,
\begin{equation}
Z(\lambda) =\sum_{N_2} \mathrm{e}^{-\lambda N_2} \tilde{Z}(N_2).
\label{}
\end{equation}
We use a Monte Carlo Markov chain (MCMC) algorithm\footnote{For details on implementing MCMC for two-dimensional CDT, see e.g.\ \cite{vdduin, DuinCode}.}
to generate sequences of independent CDT configurations, which allows
us to approximate the expectation values $\langle \beta_i\rangle$ of the Betti numbers 
\begin{figure}[t]
\centering
\scalebox{0.7}{\includegraphics[-30,-20][60,60]{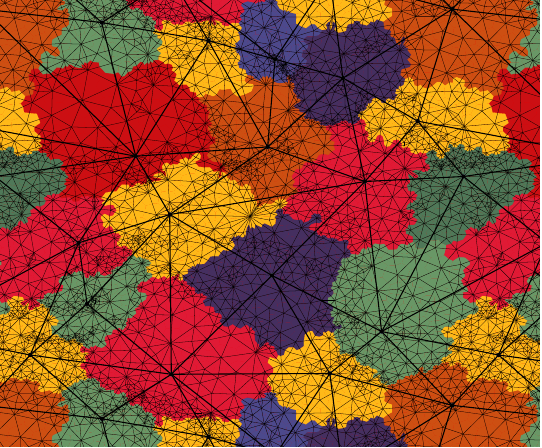}}
\hfill
\includegraphics[width=0.45\textwidth]{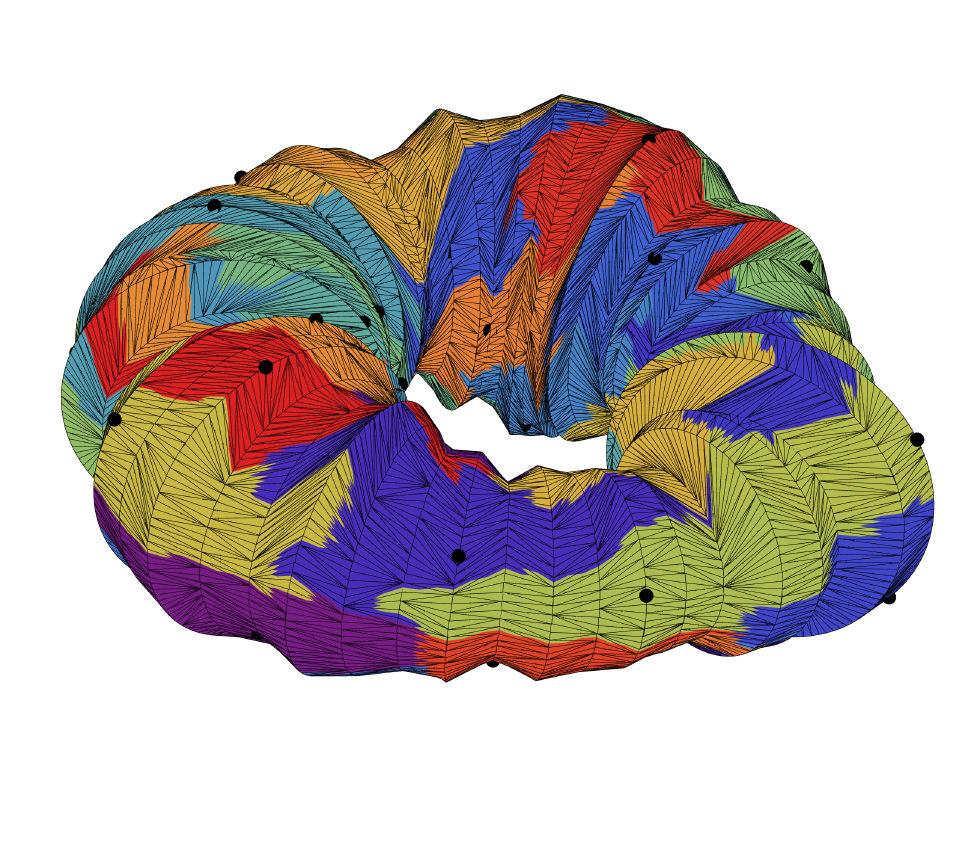}
\caption{Voronoi decompositions of typical CDT configurations on a two-torus at resolution scale $\delta=8$, in a planar representation with 
$N_2=5.040$ and $t_\mathrm{tot}=35$ (left), and an embedding in $\mathbb{R}^3$ with $N_2=10.000$ and $t_\mathrm{tot}=52$ (right).
Thin black lines are those of the original triangulation $T$. In the figure on the left, the edges of the dual Delaunay triangulation $T_\delta$ have been
added as thick black lines.}
\label{fig:torusdelta8}
\end{figure}
as the lattice volume is increased systematically.
The typical number of measurements ranged from around $410k$ for the smallest volume
$N_2=50k$ to around $40k$ for the largest volume $N_2=400k$.\footnote{For the intermediate volumes $N_2=100k,150k,200k,250k,300k$ and $350k$, the typical
number of measurements was on the order of $125k,65k,70k,45k,60k$ and $55k$ respectively. 
} 
We perform 200 sweeps between measurements, where each sweep consists of $N_2$ attempted Monte Carlo moves. Given the relatively high acceptance rate achieved in our simulations, this means that on average several Pachner moves are performed per simplex. We can therefore safely assume that our measurements are uncorrelated.

Fig.\ \ref{fig:torusdelta8} illustrates the nature of typical CDT configurations with their coloured Voronoi cells after coarse-graining with resolution $\delta=8$, in a planar
representation analogous to Fig.\ \ref{fig:voro} (left), and superimposed on the original torus in a three-dimensional embedding (right).
Each cell is associated with a vertex from the evenly spread sample ${\cal S}_8$ (indicated by a fat black dot), whose creation was described in Sec.\ \ref{sec:vertex}.
Although the original triangulation $T$ -- still visible in both representations -- is locally curved, the coarse-graining procedure contains random elements,
and the representations are not strictly isometric, 
the resulting patterns of Voronoi cells are still fairly regular. All depicted cells have disc topology and a roundish shape, unlike what we will meet in
Euclidean quantum gravity in Sec.\ \ref{sec:eucl} below.

\begin{figure}[t]
\centering
\includegraphics[width=0.49\textwidth]{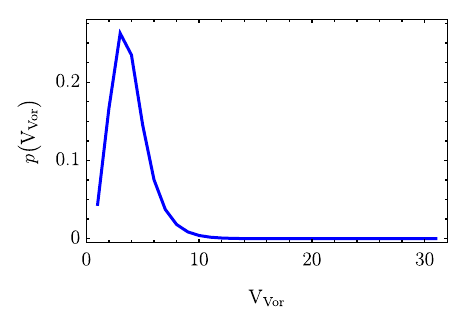}
\hfill
\includegraphics[width=0.49\textwidth]{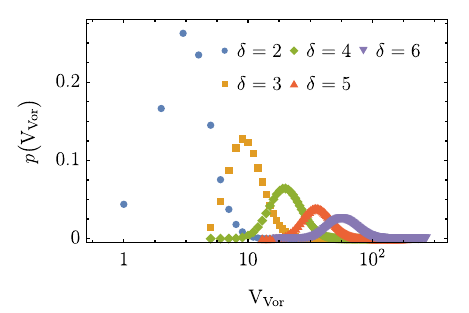}
\caption{Normalized distribution $p(V_\mathrm{Vor})$ of the volume $V_\mathrm{Vor}$ of individual Voronoi cells in units of vertices of the original triangulation $T$ for CDT configurations
of volume $N_2=200k$ and time extension $t_\mathrm{tot}=179$, for $\delta=2$ (left) and for the range $\delta\in [2,6]$ (right). In this and other data plots below, the size of error bars is smaller than the dot size. 
}
\label{fig:CDT_50k_Vor}
\end{figure}

For a more detailed understanding of our coarse-graining procedure, we have monitored the volumes of the Voronoi cells at resolution $\delta$ in terms of
the numbers of vertices of the triangulation $T$ they contain. 
As can be seen from Fig.\ \ref{fig:CDT_50k_Vor}, our algorithm has the property of creating cells of approximately equal size, with volume distributions that
remain well peaked even for $\delta >2$. Across a range of volumes $N_2$ and time extensions $t_\mathrm{tot}$ we have investigated, these distributions are
essentially unchanged. 

\begin{figure}[t!]
\centering
\includegraphics[width=0.49\textwidth]{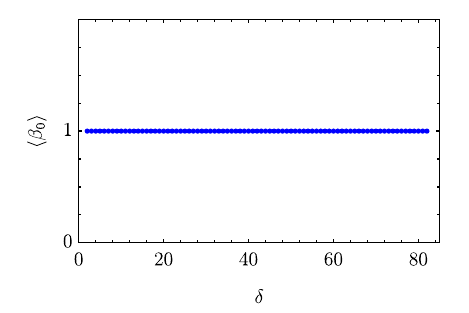}
\hfill
\includegraphics[width=0.49\textwidth]{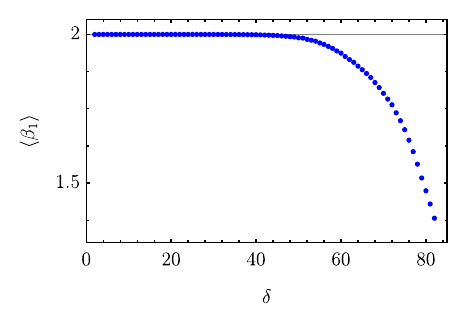}\\
\centering
\includegraphics[width=0.49\textwidth]{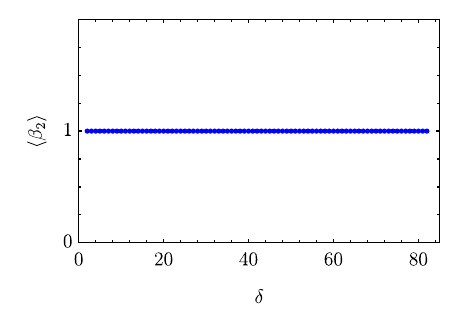}
\caption{Expectation values of the Betti numbers $\beta_0$, $\beta_1$ and $\beta_2$ in 2D Lorentzian quantum gravity
with $N_2=200k$ and $t_\mathrm{tot}=179$, as a function of the resolution $\delta\in [2,82]$.
The straight line at $\langle \beta_1\rangle=2$ is included for comparison.
}
\label{fig:betti_cdt_all}
\end{figure}

We now turn our attention to the expectation values $\langle \beta_i (\delta)\rangle$ of the Betti numbers. Fig.\ \ref{fig:betti_cdt_all} shows the measurement results
for the CDT ensemble with volume $N_2=200k$ and time extension $t_\mathrm{tot}=179$, in the range $\delta\in [2,82]$. 
The original triangulations $T$, which correspond to $\delta=1$, i.e.\ the case without any coarse-graining, by construction have the topology of a torus,
with $\beta_0=1$ counting its single connected component, $\beta_1=2$ the loops along the two directions of the torus, and $\beta_2=1$ the
single two-dimensional hole or ``cavity" enclosed by the torus.
We see that throughout most or all of the observed $\delta$-range, the expectation values
for the coarse-grained geometries $T_\delta$ reproduce these values within error bars. In other words, the effective homology coincides with the original homology
even for very large coarse-graining and in this sense does not reveal any new local structure. 

Only when the resolution $\delta$ reaches a scale where 
the nontrivial global topology of the torus geometries comes into play, the average Betti numbers $\langle\beta_1\rangle$ and $\langle\beta_2\rangle$ can be affected. This happens because
the diameter of a Voronoi cell can become large enough to wrap around one or both of the compact torus directions. This results in global pinchings of the geometry
as described in Sec.\ \ref{sec:delaunay}, where the pinchings now take place at noncontractible loops of the original triangulation. By contrast, the expectation value $\langle\beta_0\rangle$ cannot be affected,
since by construction no part of the Delaunay triangulation detaches during a pinching. 

\begin{figure}[t]
    \centering
    \vspace{3mm}
    \includegraphics[width=0.4\linewidth]{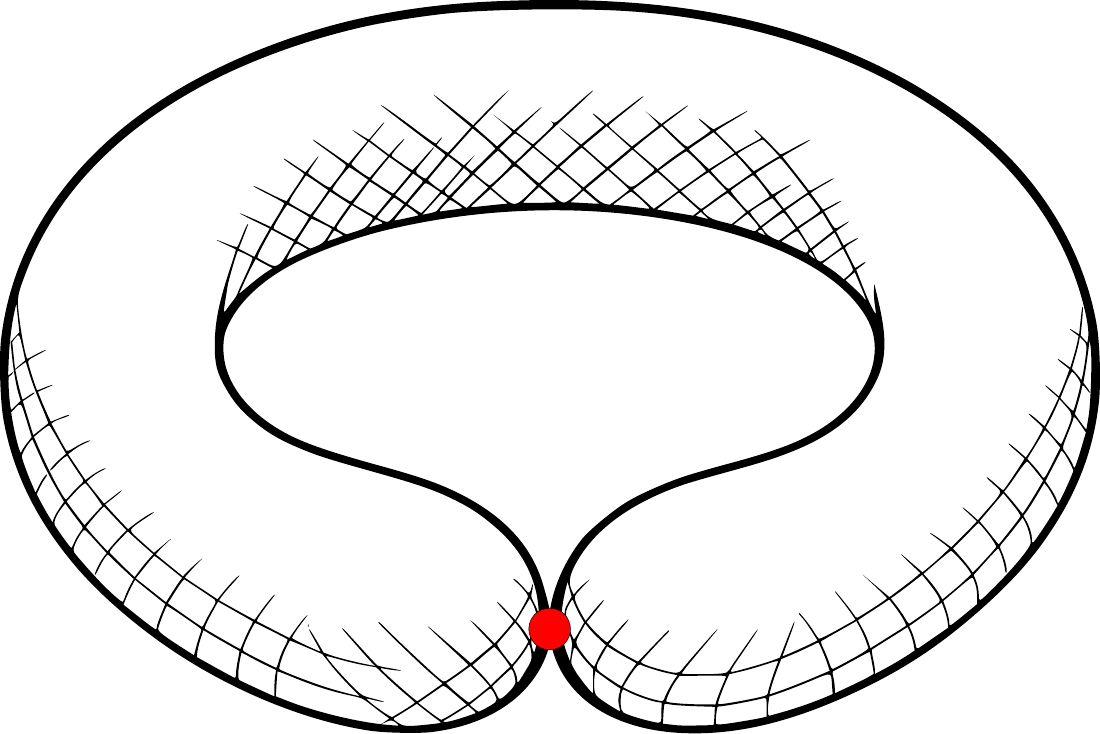}
    \vspace{3mm}
    \caption{The presence of a single global pinching vertex (red) lowers the Betti number $\beta_1$ of a toroidal Delaunay triangulation to 1.}
    \label{fig:pinched-torus}
\end{figure}

The onset of these global features in the behaviour of the Betti numbers depends both on the time extension $t_\mathrm{tot}$ and the spatial extension of the toroidal configurations. Because the size 
$\ell(t)$ of the spatial universe is subject to large quantum fluctuations (cf.\ Fig.\ \ref{fig:cdt-history}), the global effect on observables of wrapping around the compact spatial direction becomes noticeable 
on length scales much smaller than the average spatial extension $\bar{\ell}$, as has been
observed in previous studies of other observables in 2D CDT quantum gravity \cite{Brunekreef2021,correl}.
This is also what we have found in the present study. 

For the parameter choices $N_2=200k$ and $t_\mathrm{tot}=179$ of the data plots shown in Fig.\ \ref{fig:betti_cdt_all}, a nontrivial wrapping of Voronoi cells in the time direction cannot occur,
since the maximal diameter of such a cell is of the order of $2\delta$, which in the $\delta$-range probed is always smaller than $t_\mathrm{tot}$. It implies that the deviation from constancy in $\langle\beta_1\rangle$ for $\delta\gtrsim 38$ 
observed in Fig.\ \ref{fig:betti_cdt_all} is due to nontrivial global effects in the spatial direction. 
The leading contribution to the decrease of this expectation value is from configurations with 
a single global pinching vertex associated with contracting a spatial loop of winding number 1.
Namely, for a given Delaunay triangulation $T_\delta$, $\beta_1$ drops from 2 to 1 in the presence of a single global pinching,
since one of the noncontractible loops of the torus disappears, while $\beta_2$ remains unaffected, since there still is a two-dimensional cavity present. This is illustrated schematically by Fig.\ \ref{fig:pinched-torus}. 

There are additional, smaller effects related to multiple pinchings, all of which occur with higher frequency as
$\delta$ grows. For example, introducing a second pinching vertex in the configuration of Fig.\ \ref{fig:pinched-torus}, associated with the same direction but in a different location along the torus, will result in the appearance of two spherical cavities or ``bubbles", with associated Betti number $\beta_2=2$. Correspondingly, additional pinchings can increase the number of bubbles and therefore $\beta_2$ even further. 
All of these lead to an increase in $\langle\beta_2\rangle$ as $\delta$ becomes larger. 
However, in the measurement of $\langle\beta_2\rangle$ reported in Fig.\ \ref{fig:betti_cdt_all}
their occurrence is so rare that it does not lead to any appreciable deviation from constancy in the entire
$\delta$-range considered.

\begin{figure}[h!]
\centering
\includegraphics[width=0.45\textwidth]{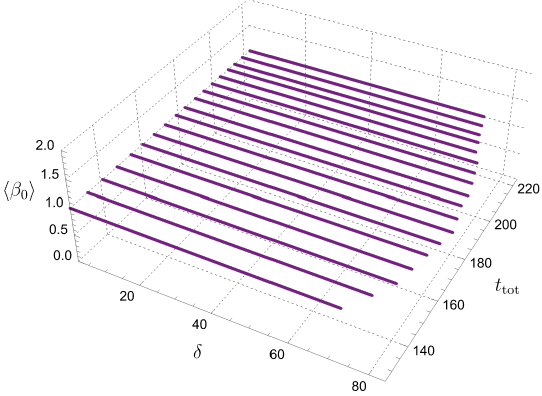}
\hfill
\includegraphics[width=0.45\textwidth]{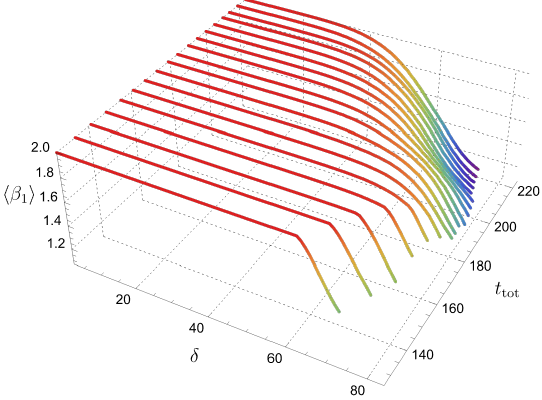}\\
\centering
\includegraphics[width=0.45\textwidth]{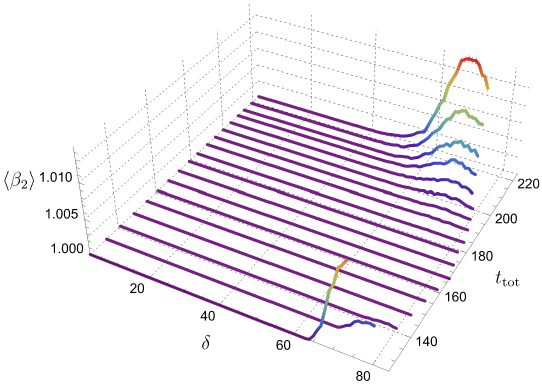}\\[6pt]
\caption{Expectation values of the Betti numbers $\beta_0$, $\beta_1$ and $\beta_2$ in 2D Lorentzian quantum gravity 
with $N_2=200k$, as a function of the resolution $\delta$ and the time extension $t_\mathrm{tot}$.
}
\label{fig:CDT_200k_Bettis}
\end{figure}

\begin{figure}[h!]
\centering
\begin{tabular}{@{}c@{}}
\includegraphics[width=0.55\textwidth]{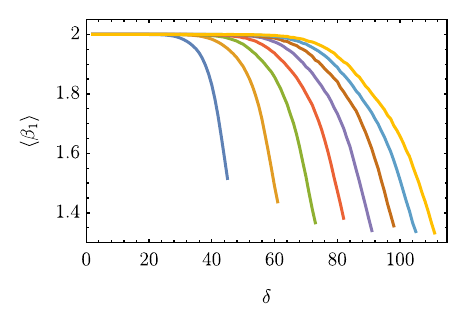}
\end{tabular}
\vspace{\floatsep}
\begin{tabular}{@{}c@{}}
\vspace{23pt}
\includegraphics[width=0.14\textwidth]{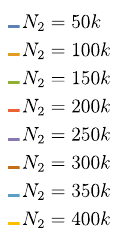}
\end{tabular}
\vspace{-18pt}
\caption{Expectation value $\langle\beta_1\rangle$ of the first Betti number in 2D Lorentzian quantum gravity as a function of the volume $N_2$, for a fixed ratio $r=0.16$.}
\label{fig:CDT_Bettis_vol}
\end{figure}

To study these effects more systematically, we have performed series of measurements of the Betti numbers at fixed volume $N_2=200k$, where in addition to
$\delta$ we also vary the time extension, in the range $t_\mathrm{tot}\in [126,224]$, see Fig.\ \ref{fig:CDT_200k_Bettis}.  
Note that for a given triangulation $T$, whenever the resulting Delaunay triangulation no longer contains any triangles we stop the coarse-graining and discard this
configuration, independent of the values of its Betti numbers. It implies that the number of measurements we average over varies as a function of $\delta$. 
We stop plotting a curve for given $t_\mathrm{tot}$ when more than 5\% of measurements are discarded this way. We have verified on a test sample that this 
procedure captures the properties of the coarse-grained triangulations correctly, while allowing us to access sufficiently large $\delta$ to exhibit the nontrivial
behaviour of the Betti numbers.

The expectation value of $\beta_0$, shown in Fig.\ \ref{fig:CDT_200k_Bettis},
remains at the constant value 1 throughout, as anticipated. The location of the
characteristic drop of the expectation value $\langle\beta_1 (\delta)\rangle$
turns out to depend on the time extension $t_\mathrm{tot}$. For small
$t_\mathrm{tot}$, it is primarily caused by pinchings along the time direction;
these become rarer for increasing time extension, pushing the drop-off to larger
values of $\delta$. Since a bigger $t_\mathrm{tot}$ at constant two-volume
implies a smaller spatial extension, pinchings along the spatial direction start
appearing, leading to a monotonic decrease in the location of the drop-off
beyond $t_\mathrm{tot}\approx 155$. 
A complementary set of measurements of $\langle \beta_1\rangle$, for increasing volume
but a fixed ratio $r$ of the time and spatial extensions is given in Fig.\ \ref{fig:CDT_Bettis_vol}. 
Again we observe that the $\delta$-interval where the expectation value is compatible with 2 grows
as the volume becomes larger, roughly speaking $\propto\sqrt{N_2}$, as we have checked. 
The chosen ratio $r=0.16$ is such that the effect comes from
global pinchings of the spatial direction only.
Lastly, as explained above, the Betti number
$\beta_2$ is sensitive to the occurrence of multiple pinchings.
Such pinchings along the time direction are in evidence in the measurements of
its expectation value for small $t_\mathrm{tot}\lesssim 148$ and large $\delta$,
while the effects of multiple pinchings along the spatial direction start
showing up in the plots for the largest time extensions $t_\mathrm{tot}\gtrsim200$ considered.
Note that the absolute size of the change in $\langle\beta_2 (\delta)\rangle$ away from the classical torus value is about two orders of magnitude smaller than that of $\langle\beta_1 (\delta)\rangle$.

\subsection{Euclidean quantum gravity in D=2}
\label{sec:eucl}

We turn next to the analysis of Euclidean quantum gravity in two dimensions. This toy model of quantum gravity is known to lie in a different universality class
from the Lorentzian model considered in Sec.\ \ref{sec:lor}. The crucial difference is the absence of a causal structure for the histories that are summed over
in the regularized Euclidean path integral
\begin{equation}
Z^\mathrm{eu}(\lambda)=\sum_{T} \frac{1}{C(T)}\, \mathrm{e}^{-\lambda N_2(T)},
\label{pi2dedt} 
\end{equation}
which otherwise is completely analogous to the Wick-rotated path integral (\ref{pi2dcdt}) of the Lorentzian theory.
As already mentioned in footnote\ \ref{fn1}, the sum is taken over EDT that are elements of a slightly generalized ensemble, compared to that of simplicial manifolds, which is known to not 
\begin{figure}[t!]
\centering
\includegraphics[width=0.6\textwidth]{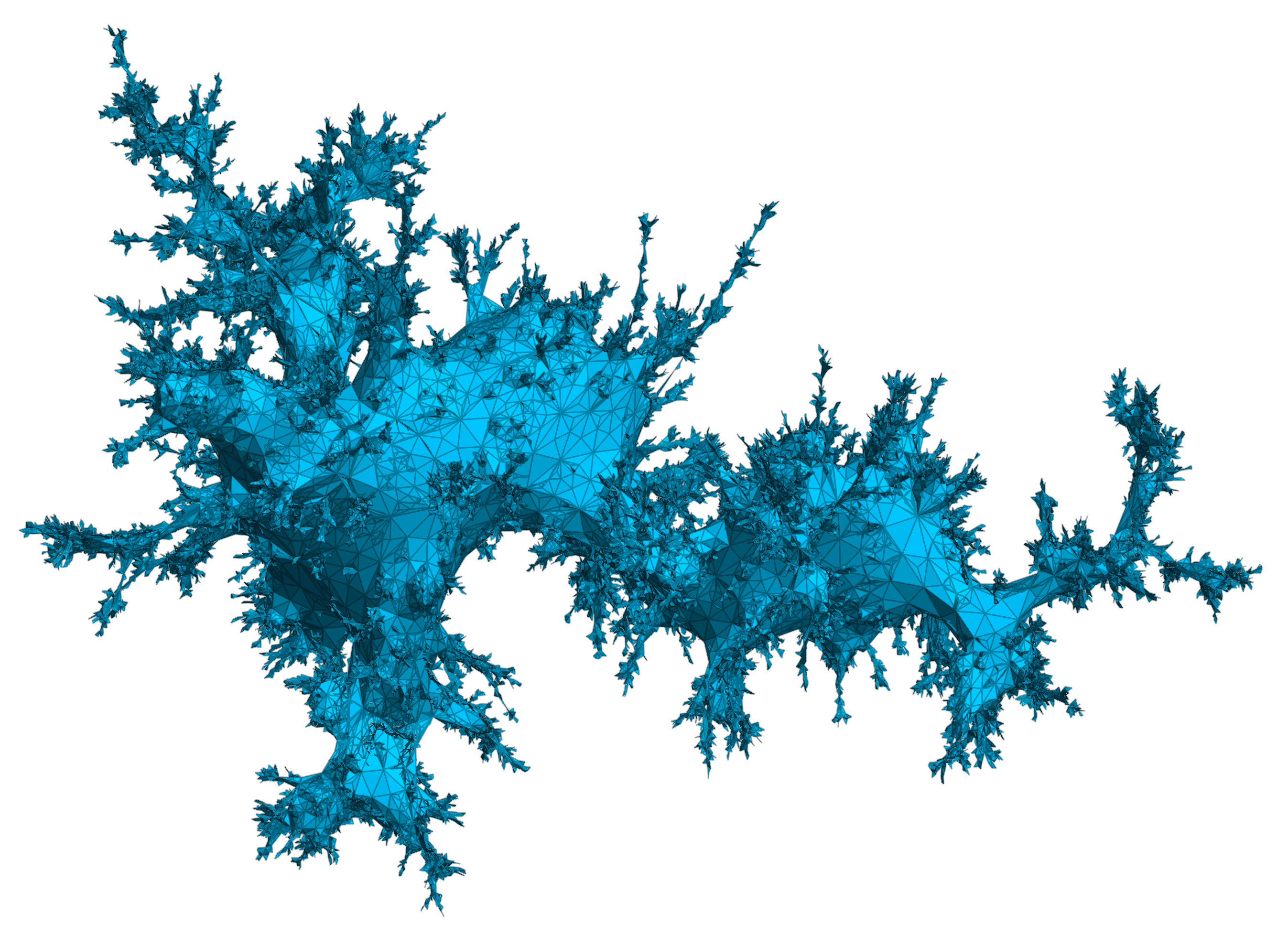}
\caption{Typical EDT configuration, with volume $N_2=100k$, depicting a Euclidean ``spacetime" of spherical topology.
}
\label{fig:edt-history}
\end{figure}
affect the continuum limit of the model. With these specifications observed, each triangulation $T$ in the sum is an arbitrary gluing of flat, equilateral triangles 
with the topology $S^2$ of a two-sphere. A typical EDT configuration is shown in Fig.\ \ref{fig:edt-history}, illustrating the well-known fractal nature of its quantum geometry.

\begin{figure}[t]
\centering
\includegraphics[width=0.6\textwidth]{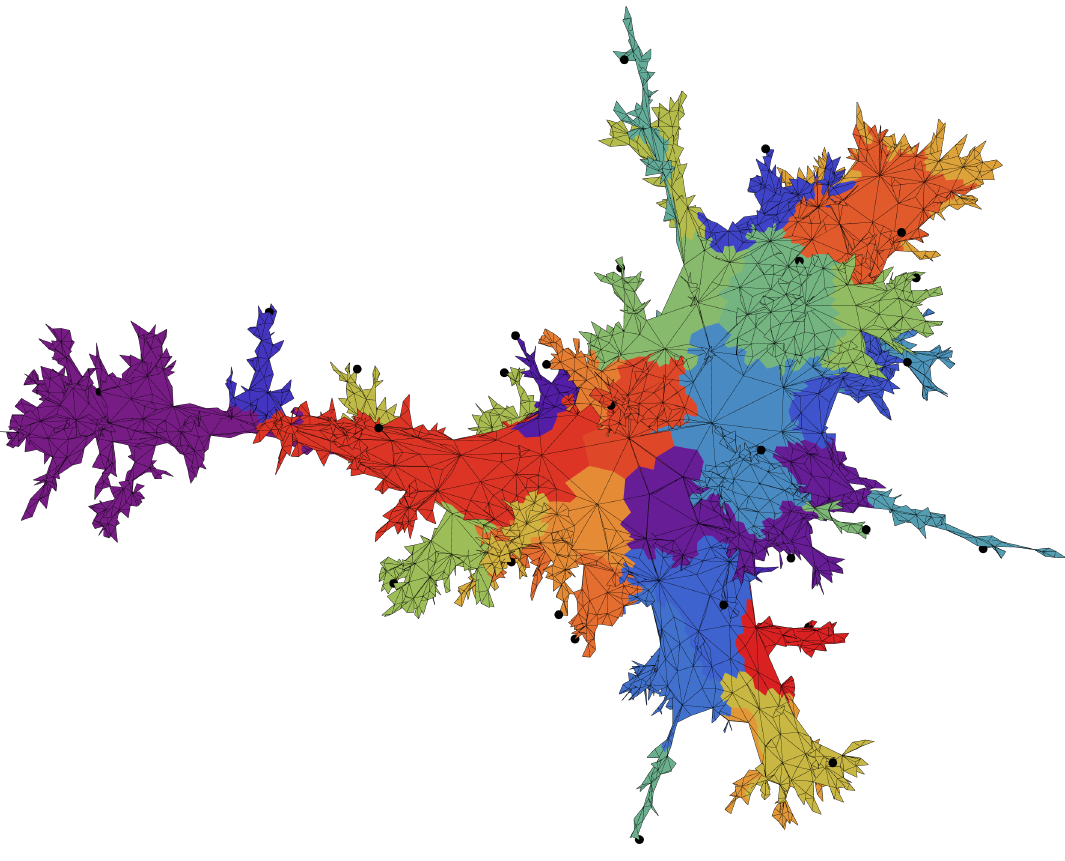}
\caption{Voronoi decomposition of a typical EDT configuration on a two-sphere at resolution scale $\delta=8$, with $N_2=10k$. 
Thin black lines are those of the original triangulation $T$.
}
\label{fig:spheredelta8}
\end{figure}
\begin{figure}[t]
\centering
\includegraphics[width=0.49\textwidth]{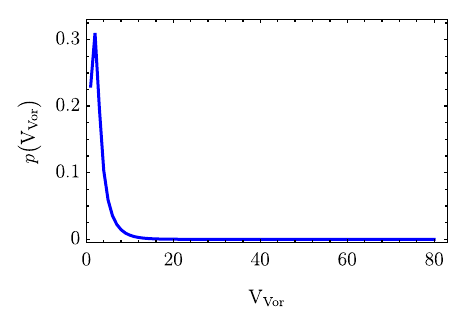}
\hfill
\includegraphics[width=0.49\textwidth]{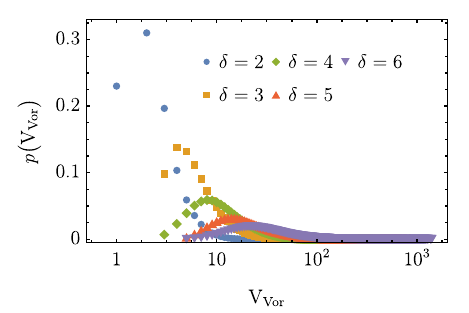}
\caption{Normalized distribution $p(V_\mathrm{Vor})$ of the volume $V_\mathrm{Vor}$ of individual Voronoi cells in units of vertices of the original triangulation $T$ for EDT configurations of volume $N_2=200k$, for $\delta=2$ (left) and for the range $\delta\in [2,6]$ (right). 
}
\label{fig:EDT_50k_Vor}
\end{figure}

We have again computed the expectation values of the Betti numbers, using the analogue of expression (\ref{expfix}) for the Euclidean fixed-volume ensemble, for
eight equally spaced values $N_2 \in [50k,400k]$. For each volume, we have performed several hundred thousand measurements on independent configurations.\footnote{For
the volumes $N_2=50k$, $100k$, $150k$, $200k$, $250k$, $300k$, $350k$ and $400k$ we used $344k$, $322k$, $568k$, $552k$, $489k$, $516k$, $440k$ and $459k$ configurations.}
The EDT configurations are generated by a direct Monte Carlo sampling method, where each sampled configuration is automatically independent.\footnote{The direct sampling uses a bijection between tadpole-free triangulations and random Kreweras excursions\cite{kreweras}, which can easily be sampled directly. The implementation is available at \cite{DuinCode}.}
Fig.\ \ref{fig:spheredelta8} shows a typical EDT configuration with coloured Voronoi cells after coarse-graining 
with resolution $\delta=8$.
Each cell is associated with a vertex from the evenly spread sample ${\cal S}_8$ (indicated by a fat black dot). Due to the fractal and ``spiky" nature of the original triangulation $T$, the shapes of the cells are much less regular than in the Lorentzian case. Note also that the annuli discussed in Sec.\ \ref{sec:delaunay} are present whenever a Voronoi cell of one colour wraps around an outgrowth that continues with a cell of another colour.

\begin{figure}[t]
\centering
\includegraphics[width=0.49\textwidth]{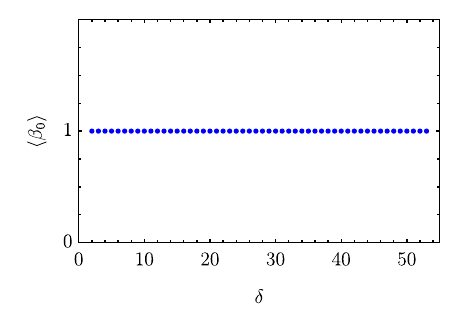}
\hfill
\includegraphics[width=0.49\textwidth]{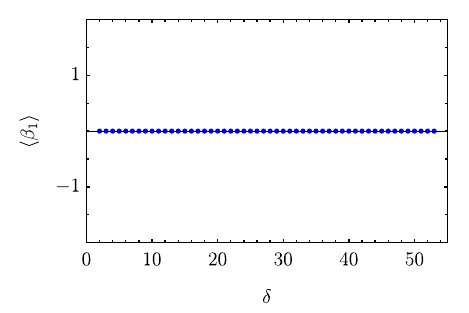}\\
\centering
\includegraphics[width=0.49\textwidth]{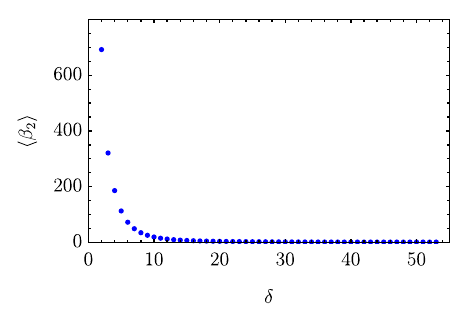}
\caption{Expectation values of the Betti numbers $\beta_0$, $\beta_1$ and $\beta_2$ in 2D Euclidean quantum gravity
with $N_2=200k$, as a function of the resolution $\delta\in [2,53]$. }
\label{fig:betti_edt_all}
\end{figure}

We have checked that our algorithm nevertheless distributes volume among the cells relatively evenly. This is illustrated by Fig.\ \ref{fig:EDT_50k_Vor}, 
which is the Euclidean analogue of Fig.\ \ref{fig:CDT_50k_Vor} above, showing the distribution of volumes of the Voronoi cells at resolution $\delta$ in terms of
the numbers of vertices of the triangulation $T$. The situation is qualitatively similar to that of the Lorentzian case, with the peaks of the distributions located at slightly lower
 values of the cell volume $V_\mathrm{Vor}$, and somewhat broader distributions for the larger $\delta$-values. 

The measurements of the expectation value of the Betti numbers for the EDT ensemble with volume $N_2=200k$ in the range $\delta\in [2,53]$
are displayed in Fig.\ \ref{fig:betti_edt_all}.
Before coarse-graining, the triangulations $T$ have the topology of a sphere,
with $\beta_0=1$ counting its single connected component, $\beta_1=0$ reflecting the absence of noncontractible loops, and $\beta_2=1$ for the
single two-dimensional hole enclosed by the sphere. 
Throughout the entire observed $\delta$-range, not only the expectation values $\langle \beta_0\rangle$ and $\langle \beta_1\rangle$ but the Betti numbers $\beta_0$ and $\beta_1$ for each individual coarse-grained geometry 
reproduce these values. The constancy of $\beta_0$ is unsurprising, since the algorithm we use preserves the connected character of the geometries.
In the present case, since the geometries
remain also simply connected, this also explains the constancy of $\beta_1$. However, pinchings can in principle happen, where the sphere after coarse-graining becomes
a set of connected spherical bubbles, like the structure sketched in Fig.\ \ref{fig:bubbles} above. This changes $\beta_2$, which counts the number of such bubbles.

The bottom plot of Fig.\ \ref{fig:betti_edt_all} for the expectation value of $\beta_2$ indicates that this is exactly what happens here and, unlike in the Lorentzian
case, is a \textit{local} phenomenon. Already for the smallest coarse-graining step $\delta=2$, the expectation value of $\beta_2$ shoots up to a maximum, then 
decreases steeply as the resolution becomes larger, and asymptotes to 1 for the largest
$\delta$-values considered.\footnote{Since we do not include configurations where all triangles have vanished, we always have $\beta_2\geq 1$.} 
\begin{figure}[t]
\centering
\includegraphics[width=0.49\textwidth]{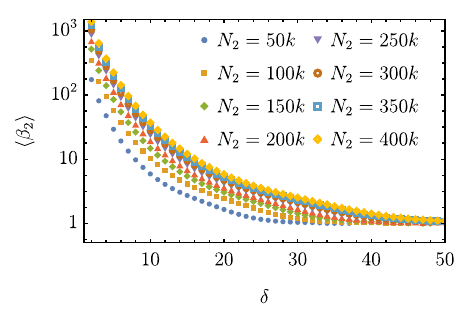}
\hfill
\includegraphics[width=0.49\textwidth]{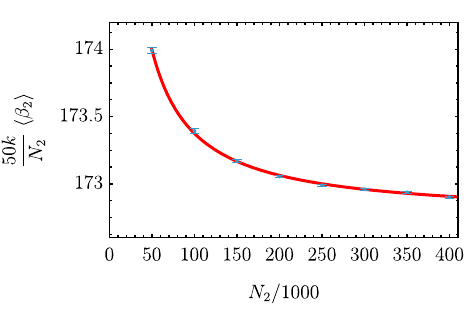}
\caption{Expectation value of the Betti number $\beta_2$ in 2D Euclidean quantum gravity 
as a function of the resolution $\delta$, for volumes $N_2\in [50k,400k]$ (left). 
The same expectation value, for $\delta =2$, rescaled by the volume, and plotted as a function of the volume $N_2$. 
The red line is the best fit to $\tfrac{50k}{N_2}\langle \beta_2\rangle=A+B/N_2$, for fitting constants $A$ and $B$ (right).}
\label{fig:EDT_Betti_2_rescs}
\end{figure}
A similar behaviour can be observed for a range of volumes $N_2\in[50k,400k]$, where correspondingly more bubbles appear, as shown in Fig.\ \ref{fig:EDT_Betti_2_rescs}, left. The locality of this ``bubble generation" is underscored by the
fact that for $\delta=2$, 
\begin{figure}[ht!]
\centering
\includegraphics[width=.93\linewidth]{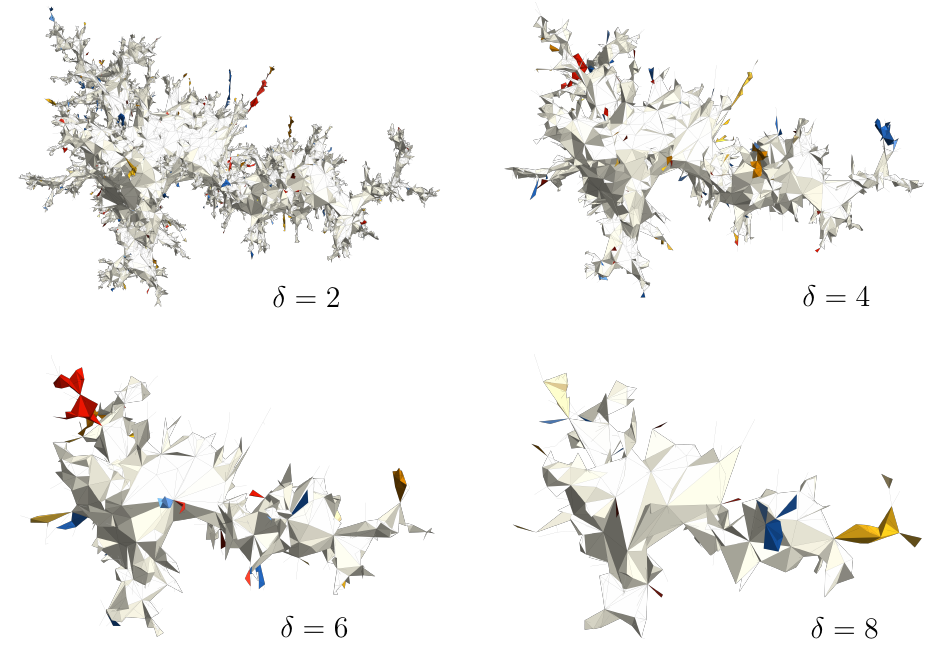}
\caption{Coarse-grainings of the spherical EDT configuration of Fig.\ \ref{fig:edt-history} with initial volume $N_2=100k$, 
at resolution $\delta=2,4,6$ and 8, showing a large mother universe (white) and
bubbles that appear due to pinching (coloured).
}
\label{fig:EDTseries}
\end{figure}
for which we have the best bubble statistics, the average Betti number $\langle \beta_2\rangle$ for large $N_2$ scales approximately
linearly with the volume, as illustrated by Fig.\ \ref{fig:EDT_Betti_2_rescs}, right.

\begin{figure}[t]
\centering
\includegraphics[width=.82\linewidth]{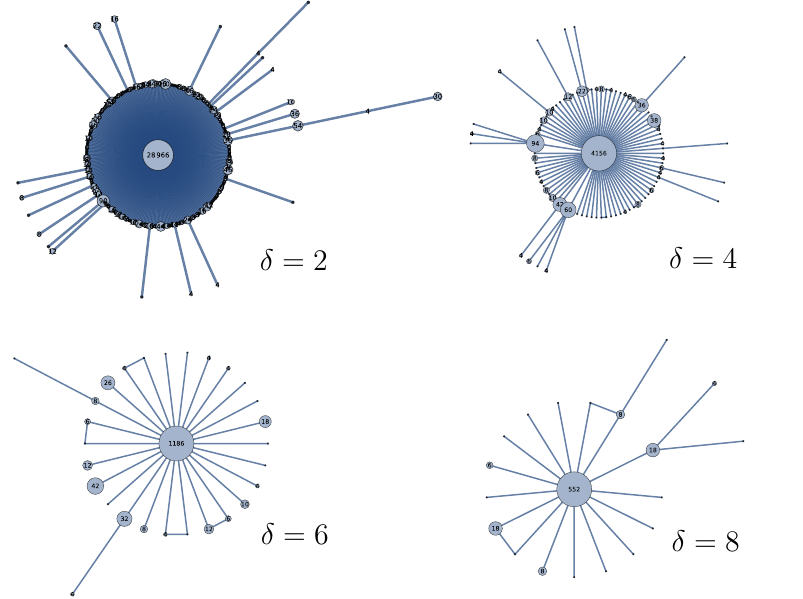}
\caption{Bubble diagrams of the four coarse-grained Delaunay triangulations depicted in Fig.\ \ref{fig:EDTseries}, with discs associated to bubbles, and bubble volumes given in terms of coarse-grained triangle units, as described in the main text.}
\label{fig:EDTbubbleseries}
\end{figure}

The generation of these spherical bubbles is illustrated further by the sequence of Delaunay triangulations 
$T_\delta$, for $\delta=2,4,6$ and 8, depicted in Fig.\ \ref{fig:EDTseries},
which are coarse-grained versions of the triangulation of Fig.\ \ref{fig:edt-history}. At every iteration there is a ``mother universe", drawn in white, which
contains most of the volume, and smaller, coloured outgrowths or bubbles that are connected to the mother universe or to other bubbles by pinchings or loose edges. To convey a more quantitative understanding of the number and sizes of the bubbles, we present another diagrammatic representation of the same four configurations in Fig.\ \ref{fig:EDTbubbleseries}. Individual spherical components or
bubbles are represented by discs whose radius is proportional to $\log\!\big( \tilde{N}_2/2 \big)$, where $\tilde{N}_2$ is the number of coarse-grained triangles in the bubble, which is also displayed alongside the disc. (The label for bubbles with $\tilde{N}_2 = 2$ is suppressed to avoid clutter.) A line is drawn between bubbles if they are connected by a 
pinching vertex or by one or more loose edges in the coarse-grained triangulation $T_\delta$.\footnote{Note that this leads to closed
loops in the bubble diagram whenever there is a triple or higher-order meeting point between bubbles in $T_\delta$. In Fig.\ \ref{fig:EDTbubbleseries} this happens in the diagrams for $\delta=6$ and $\delta=8$. It does \textit{not} imply that $T_\delta$ itself contains noncontractible loops (which it does not).} 
All diagrams exhibit the presence of a mother universe at the centre, where the by far largest fraction of the volume is located, with a
first generation of much smaller neighbouring bubbles, a much sparser second generation of small bubbles, and
occasional instances of bubbles of a higher generation (in the examples shown, these are only present
in the diagram for $\delta=2$).

Unlike what we found in 2D Lorentzian quantum gravity, the nontrivial aspects of the effective
homology in the Euclidean case are not just global, but also local\footnote{Strictly speaking, in 2D quantum gravity without topology change there is no distinction between local and global, since any length scale is set by a cosmological constant $\Lambda$, the continuum counterpart of the coupling $\lambda$ appearing in the bare actions in (\ref{pi2dcdt}) and (\ref{pi2dedt}). However, in 2D CDT, by choosing suitable combinations of the two-volume $N_2$ and the time extension $t_\textrm{tot}$, we can create a ``bulk regime" largely free of the effects of global winding numbers, as we have seen.}, in the sense that the
expectation value $\langle \beta_2(\delta)\rangle$ behaves nonclassically for all values of $\delta$.
This behaviour is related to the well-known fractal structure of 2D Euclidean quantum gravity. 
It can be characterized by the presence of so-called minimal-neck baby universes (''minbus"), parts of the two-dimensional 
geometry that are connected to a mother universe by minimal necks, consisting of closed loops of
three lattice links \cite{Jain1992,Ambjorn1993}. Although our construction does not exactly identify such minbus, we show in \cite{letter} that the bubble structure we find is close enough to establish 
(for $\delta=2$) a quantitative relation with 
the string susceptibility, a scaling parameter that in these references is extracted by measuring
the statistical distribution of minbu sizes.
Our bubble diagrams of Fig.\ \ref{fig:EDTbubbleseries} suggest that the fractal-like, hierarchical structure of a
mother universe and subsequent generations of smaller bubbles is preserved during coarse-graining, at least at a qualitative level.

\section{Conclusions and Outlook}
\label{sec:concl}

By explicit construction, we have shown that the tool of \textit{effective homology} can be used to characterize the quantum geometry in two-dimensional models of quantum gravity, 
defined nonperturbatively as continuum limits of dynamically triangulated lattice theories. The key idea is to construct a set of observables  
describing local metric properties of the quantum space(-time) by using the powerful machinery of TDA after applying a local coarse-graining algorithm to the path integral configurations.
A coarse-graining of resolution $\delta$ produces a (generalized) triangulation with typical edge length $\delta$ (an integer in original lattice units): $\delta=1$ reproduces the original triangulation, $\delta=2$ results in a triangulation with edge length 2 (and correspondingly fewer triangles), and so forth.
Although the original triangulations all have the same, fixed topology (of a torus in the Lorentzian and a sphere in the Euclidean case), this need not be the case after coarse-graining,
since the latter by construction does not resolve substructures of linear size smaller than $\delta$.  

The concrete observables we measured on lattices of volume $N_2\leq 400k$ were the Betti numbers $\beta_i(\delta)$, $i=0,1,2$, as a function of the coarse-graining scale. 
The only nontrivial behaviour of their expectation values we found in the Lorentzian model comes from large $\delta$, where the cells of the coarse-grained
Voronoi decomposition become sufficiently large to completely wrap around one or both of the torus directions, leading to singular pinchings of the associated Delaunay 
triangulation that affect both $\langle\beta_1\rangle$ and $\langle\beta_2\rangle$ (Fig.\ \ref{fig:CDT_200k_Bettis}). A similar
pinching mechanism is also present in the Euclidean model, but already at a local scale, which affects the expectation value $\langle\beta_2\rangle$ already at the smallest nontrivial
coarse-graining $\delta=2$ (Fig.\ \ref{fig:betti_edt_all}). 

Returning to the theme of symmetry we introduced in Sec.\ \ref{sec:intro}, since the Lorentzian and Euclidean quantum gravity models in 2D do not have nontrivial classical limits, 
we cannot examine the issue of \textit{recovering} any symmetries, but we can still ask whether their quantum geometries may support continuous isometries of some kind. For
the Lorentzian model, our analysis has not found any obstructions, at least not of a local kind and when staying away from length scales where global pinchings can occur. By contrast, the bubble structure of Euclidean quantum gravity presents a clear obstacle to the existence of such symmetries. 

Our investigation demonstrates the feasibility of the concept of effective homology and its technical implementation, for nontrivial quantum gravity models of either spacetime signature in two dimensions. We have rediscovered
the well-known fractal structure of Euclidean Liouville gravity, consisting of mother and baby universes, but otherwise have not discovered any new features. Given the 
well-documented and relatively straightforward properties of both the Lorentzian and Euclidean models, this was not to be expected either. However, our
explicit set-up did not rely in an essential way on working in two dimensions, which opens the way to its application in four dimensions and Lorentzian signature, 
which is our main target and motivation. 

The higher-dimensional case will be richer and inevitably more complex, with 
more freedom of how to handle topology changes during coarse-graining, generalizing the ``pinching" that was the main feature in two dimensions.
Any choices should be guided by what is expected to be important for phenomenologically interesting observables, including those related to matter coupling. 
Also, computational efficiency and lattice size, which hardly played a role in the current study, will become factors to be considered in $D>2$. 
The 4D application will break genuinely new ground, given our current, limited understanding of its local quantum geometry \cite{ency,Loll2025} and how to characterize it in terms of
suitable observables. Is it really a quantum foam on short scales, and does it support local or global symmetries when coarse-grained suitably? -- We look forward to investigating the physical case of 4D Lorentzian quantum gravity in the near future.

\subsubsection*{Acknowledgments} 
The contribution of JvdD and RL is supported by a grant in the Open Competition ENW-M Program of the Dutch Research Council (NWO), with 
grant ID \url{https://doi.org/10.61686/IBVAP30787}. The research of MS was supported by a Radboud Excellence fellowship from Radboud University in Nijmegen, Netherlands, and by an NWO Veni grant under grant ID  [\url{https://doi.org/10.61686/SUPEH07195}].
JvdD and RL thank the Perimeter Institute for hospitality.
This research was supported in part by Perimeter Institute for Theoretical Physics. Research at Perimeter Institute is supported by the 
Government of Canada through the Department of Innovation, Science and Economic Development and by the Province of Ontario through
the Ministry of Colleges and Universities. 

\appendix
\vspace{0.5cm}

\section*{Appendix A}
\label{app:A}

In this appendix, we provide some details about how we construct the Voronoi decomposition of a triangulation $T$ 
whose dual is a coarse-grained Delaunay triangulation $T_\delta$ of a given resolution $\delta$. 
Our starting point is the original triangulation $T$, where each triangle of $T$ has already been coloured and decorated according to the prescription outlined 
in Sec.\ \ref{sec:voronoi} and illustrated in Fig.\ \ref{fig:threetri}. We separately keep track of type-1 and type-2 triangles, to make sure that during the
construction of the Voronoi decomposition every piece of boundary segment and every triple intersection of such segments has been taken into account.

\begin{figure}[h!]
\newcommand{\triangleplot}[1]{
    \begin{minipage}{0.3\linewidth}
    \includegraphics[width=\linewidth]{#1}
    \end{minipage}
}
\centering
\triangleplot{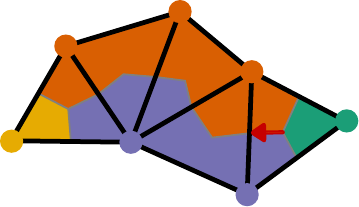}
\hfill
\triangleplot{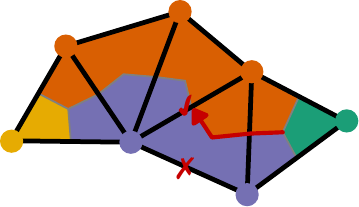}
\hfill
\triangleplot{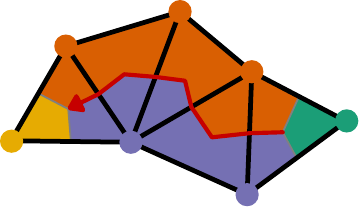}
\caption{Moving through a sequence of triangles to identify a boundary segment between two triple points belonging to the Voronoi decomposition.
}
\label{fig:tripath}
\end{figure}

Our algorithm proceeds by considering each type-2 triangle in turn, and establishing the three nearest triple points that can be reached from the
triple point at the centre of this triangle by following boundaries between differently coloured regions on the triangulation $T$. This determines the
three boundary segments of the Voronoi decomposition meeting at the initial triple point. 

The process of finding a
nearest neighbour\footnote{We use a data structure where together with each triangle we store the labels of its neighbouring triangles, 
such that finding a neighbour is simply a
look-up in constant time.} is illustrated by Fig.\ \ref{fig:tripath}. Starting from a triple vertex (left figure), we choose to walk along one of the three boundaries to 
the centre of the next triangle, in this case by crossing the triangle edge between the violet and the orange vertex of the triangle. There are then two
possibilities: if the second triangle is of type 2, we have found the other triple point where the boundary segment ends; we then go back to the first
triangle and explore the next direction, say, anticlockwise. If the second triangle is of type 1 (as is the case here, central figure), 
we next cross the triangle edge which again has a violet vertex on the left (in the direction indicated by
the arrow) and an orange vertex on the right. Such an edge always exists, since the third vertex of the triangle has to be either violet or orange. 
We then repeat the same step by continuing to the next triangle, always crossing the triangle edge with the violet vertex on the left and the orange one  
on the right, until the process ends when we encounter a type-2 triangle with a triple point (right figure). Each time we traverse a type-1 triangle, we remove it
from the list, to keep track of which such triangles have been visited.  

In this manner we exhaust all type-2 triangles. It can then happen that not all type-1 triangles have been visited yet. They must be parts of
closed boundary segments without any triple points, like the ones we met earlier in Sec.\ \ref{sec:propdel} and Fig.\ \ref{fig:loose3d2d}. 
One proceeds by picking such a triangle from the list of remaining type-1 triangles and following the boundary between the differently coloured regions
by traversing subsequent triangles, analogous to what is shown in Fig.\ \ref{fig:tripath}. Eventually this path must close on itself. This process is repeated until
no type-1 triangles are left. In this way we guarantee that all boundaries have been visited. The result is the Voronoi decomposition, whose cells, boundary
segments and triple points we associate with the dual vertices, edges and triangles respectively of the Delaunay triangulation, as described in Sec.\ \ref{sec:delaunay}.

\section*{Appendix B}
\label{app:B}

This appendix provides some details of our procedure to adjust a Delaunay triangulation such that it provides a well-defined input to GUDHI's 
computation of Betti numbers, without changing its homology. 
This input requires that each $p$-simplex is given in the form of a list $\{v_0,v_1,\dots,v_p\}$ of $p+1$ vertices. 
It cannot handle a situation where distinct edges share the same end points $\{u,v\}$ or distinct triangles share the same
corner points $\{u,v,w\}$. Both of these cases can arise during the coarse-graining procedure of Sec.\ \ref{sec:coarse}, as was explained there.

\begin{figure}[h]
\centering
\includegraphics[width=1\textwidth]{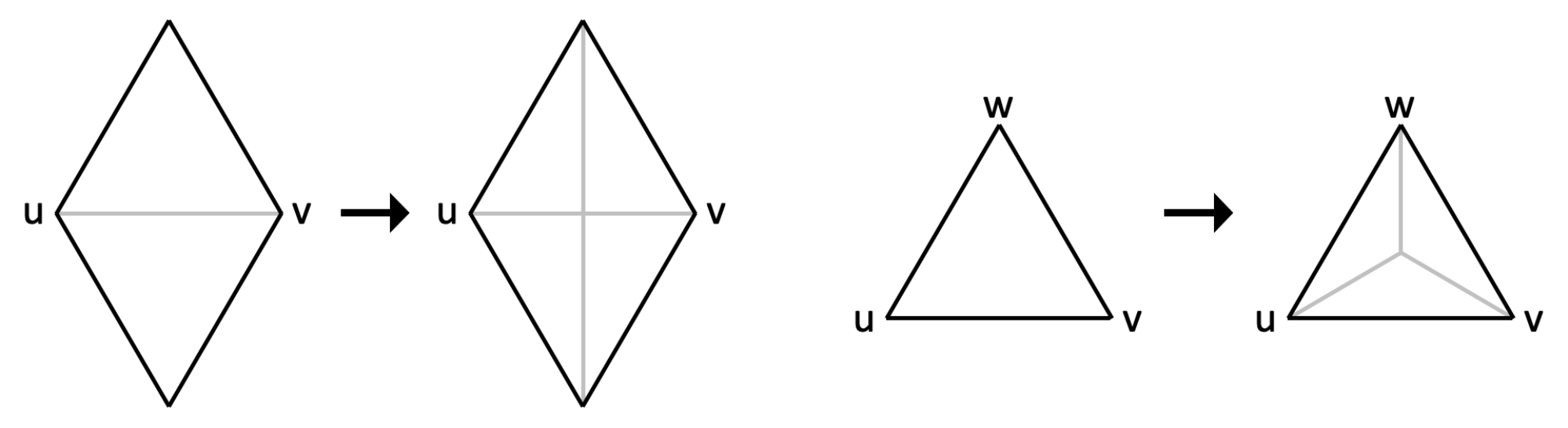}
\caption{Removing irregularities from Delaunay triangulations: multiple edges sharing the same two vertices $u$ and $v$ are subdivided (left); multiple triangles
sharing the same three vertices $u$, $v$ and $w$ are subdivided (right).}
\label{fig:regmove}
\end{figure}

To eliminate these irregularities systematically, we proceed in two steps. First, we identify all vertex pairs of the Delaunay triangulation which are connected by more than one edge.
For each such vertex pair $\{u,v\}$, we insert an additional vertex in the middle of all but one of the edges connecting $u$ and $v$, such that there are two
edges instead of one. At the same time, the triangles on either side of the edge are subdivided accordingly, such that there are four triangles instead of two, 
as illustrated by Fig.\ \ref{fig:regmove}, left. 
(If the edge was a loose edge, there are no triangles to be subdivided.) It can in principle happen that a newly created edge is part of a loop of length two,
in which case the same procedure has to be applied to it also.

After all loops of length two have been eliminated, we identify all vertex triples of the Delaunay triangulation which are shared by more than one triangle. 
For each such triple $\{u,v,w \}$, we insert an additional vertex in the middle of all but one of the triangles with corner points $u$, $v$ and $w$, such that there are three
triangles instead of one, as illustrated by Fig.\ \ref{fig:regmove}, right. Note that the two operations depicted here coincide with local Monte Carlo moves 
implemented in two-dimensional CDT and EDT respectively (see, for example, \cite{simul,ACMbook}), and do not change the connectivity of the triangulations.
The final result is a two-dimensional simplicial complex, from which we prepare the input for GUDHI in the form of three lists:
\begin{itemize}
\item a list of vertices: $\{ v_0,v_1,\dots ,v_{N_0}\}$,
\item a list of edges: $\{ \{u_0,v_0\},\{ u_1,v_1\},\dots \}$, and
\item a list of triangles: $\{ \{u_0,v_0,w_0\}, \{ u_1,v_1,w_1\},\dots \}$.
\end{itemize}

\vspace{0.5cm}


\end{document}